\documentclass[12pt]{article}
\usepackage{adjustbox}
\usepackage{indentfirst}
\usepackage{mathrsfs}
\usepackage{amssymb}
\usepackage{amsmath}
\usepackage{bbm}
\usepackage{mathtools}
\usepackage{ascmac}
\usepackage{amsthm}
\usepackage{cmll}
\usepackage{stmaryrd}
\usepackage{graphicx}
\usepackage{caption}
\usepackage{lscape}
\usepackage{subcaption}
\usepackage{natbib}
\usepackage{setspace}
\usepackage{bm}
\usepackage{url}
\usepackage{booktabs}
\usepackage{color}
\usepackage{enumitem}
\usepackage{tabularx}
\usepackage{algorithm}
\usepackage{algpseudocode}

\usepackage{times}

\newcommand{\blind}{0}

\usepackage{titlesec}
\titleformat*{\section}{\large\bfseries}
\titleformat*{\subsection}{\it}

\setlist{nosep}

\usepackage[dvipsnames]{xcolor}
	\usepackage[pdftex]{hyperref}
	\hypersetup{colorlinks,
	linkcolor=Black,  
	citecolor=NavyBlue  
	}

\usepackage{subfiles}

\def\C{\mbox{\boldmath$C$}}

\def\x{{\bm{x}}}

\def\btheta{{\bm{\theta}}}

\def\m{\mbox{\boldmath$m$}}

\theoremstyle{definition}

\usepackage{soul}

\begin{document}

\def\spacingset#1{\renewcommand{\baselinestretch}%
{#1}\small\normalsize} \spacingset{1}

\newcommand\lucio[1]{{\textcolor{purple}{#1}}}
\newcommand\ML[1]{{\sc\textcolor{cyan}{#1}}}
\newcolumntype{C}[1]{>{\centering\let\newline\\\arraybackslash\hspace{0pt}}m{#1}}

\if0\blind
{
  \title{\bf Predictive Synthesis under Sporadic Participation: Evidence from Inflation Density Surveys}
\author{
Matthew C. Johnson\thanks{Amazon. {\scriptsize Email: \texttt{jnmmatt@amazon.com}}},\,
Matteo Luciani\thanks{Board of Governors of the Federal Reserve System. {\scriptsize Email: \texttt{matteo.luciani@frb.gov}}},\\
Minzhengxiong Zhang\thanks{Department of Statistical Science, Fox School of Business, Temple University, Philadelphia, PA 19122. {\scriptsize Email: \texttt{tuj77601@temple.edu}}}, \& Kenichiro McAlinn\thanks{Department of Statistical Science, Fox School of Business, Temple University, Philadelphia, PA 19122. {\scriptsize Emails: \texttt{kenichiro.mcalinn@temple.edu}}}
}
  \maketitle
} \fi

\if1\blind
{
  \bigskip
  \bigskip
  \bigskip
  \begin{center}
    {\LARGE\bf Predictive Synthesis under Sporadic Participation: Evidence from Inflation Density Surveys}
\end{center}
  \medskip
} \fi

\bigskip
\begin{abstract}
\noindent Central banks rely on density forecasts from professional surveys to assess inflation risks and communicate uncertainty. A central challenge in using these surveys is irregular participation: forecasters enter and exit, skip rounds, and reappear after long gaps. In the European Central Bank's Survey of Professional Forecasters, turnover and missingness vary substantially over time, causing the set of submitted predictions to change from quarter to quarter. Standard aggregation rules-- such as equal-weight pooling, renormalization after dropping missing forecasters, or ad hoc imputation-- can generate artificial jumps in combined predictions driven by panel composition rather than economic information, complicating real-time interpretation and obscuring forecaster performance.
We develop coherent Bayesian updating rules for forecast combination under sporadic participation that maintain a well-defined latent predictive state for each forecaster even when their forecast is unobserved. Rather than relying on renormalization or imputation, the combined predictive distribution is updated through the implied conditional structure of the panel. This approach isolates genuine performance differences from mechanical participation effects and yields interpretable dynamics in forecaster influence. In the ECB survey, it improves predictive accuracy relative to equal-weight benchmarks and delivers smoother and better-calibrated inflation density forecasts, particularly during periods of high turnover.
\end{abstract}

\noindent%

\vfill

\renewcommand{\thefootnote}{ } 
\footnotetext{\\ 

\noindent Disclaimer \# 1: The views expressed in this paper are those of the author and do not necessarily reflect the views and policies of the Board of Governors or the Federal Reserve System. \smallskip

\noindent Disclaimer \# 2: This publication is not related to Amazon and does not reflect the position of the company and its subsidiaries. Its contents are part of a continued
collaboration based on work done prior to joining Amazon.
} 

\renewcommand{\thefootnote}{\arabic{footnote}}

\newpage

\spacingset{1.2} 

%
%
\section{Introduction}

Surveys of inflation expectations play a central role in inflation forecasting and in the assessment of risks to price stability, and they are closely monitored by both policymakers and market participants. Major central banks, including the Federal Reserve \citep{yellen2015} and the European Central Bank \citep{PC@ECB}, routinely incorporate survey-based inflation expectations into their forecasting frameworks. Beyond point forecasts-- such as medians used in expectations-augmented Phillips curve models-- these institutions also rely on the full distribution of survey responses to assess inflation risks, as in the Federal Reserve Bank of Philadelphia's Survey of Professional Forecasters (SPF) and the European Central Bank's SPF. Given the central role of these surveys in institutional inflation forecasting, movements in estimated inflation expectation distributions should reflect genuine economic information rather than mechanical artifacts of forecast aggregation.

A basic but consequential feature of many professional forecasting surveys is that participation is highly irregular. In the European Central Bank's SPF, for example, forecasters frequently skip rounds, enter or exit the panel, and exhibit long and uneven participation gaps. When participation is sporadic, standard aggregation methods can generate artificial volatility in combined density forecasts, distorting perceived changes in uncertainty and tail risk. Existing approaches to forecast combination provide limited guidance for combining survey-based density forecasts when participation is irregular. This paper formalizes this problem within a probabilistic framework and develops a coherent solution that is theoretically grounded, computationally simple, and practically effective.

When participation is sporadic, standard operational aggregation-- pooling whatever densities are currently available after dropping missing forecasters and renormalizing weights, or filling gaps via simple imputations such as last-observation-carried-forward-- can create artificial movements in combined \emph{uncertainty} and \emph{tail risk}. These artifacts arise because the pooled object changes when the contributor set changes: the same underlying economic environment can yield different pooled dispersion simply because a high-variance or outlier forecaster enters or exits.  This is the operational problem we address: separating genuine shifts in uncertainty from artifacts of entry, exit, and intermittent participation so that dispersion and tail-risk signals remain interpretable for real-time monitoring.

We document these issues in the ECB SPF for year-ahead euro-area inflation, a setting in which participation is highly uneven. More than half of forecasters do not participate regularly, long gaps are common, and entry and exit patterns vary substantially over time. These features are especially pronounced during periods of macroeconomic stress, when both forecast dispersion and panel instability increase simultaneously. In such episodes, mechanical changes in combined forecast uncertainty are particularly difficult to distinguish from genuine shifts in perceived inflation risk.

Sporadic participation is not a peculiarity of the ECB SPF. Many institutional forecasting systems-- including the Philadelphia Fed SPF and a range of private-sector consensus platforms-- aggregate forecasts from panels with persistent entry, exit, and intermittent participation patterns that often intensify during periods of heightened uncertainty. Balanced panels are operationally infeasible in these environments, so an aggregation rule must be reliable under irregular participation rather than relying on ad hoc renormalization or long-horizon imputations. Our goal is explicitly \emph{real-time aggregation} under the information that is actually reported, while preventing administrative variation in reporting from being misread as changing uncertainty.

To address these challenges, we develop a principled, turnover-aware predictive synthesis approach that maintains a continuous latent assessment for every forecaster, even when forecasts are temporarily missing. Rather than dropping a missing forecaster, which induces discontinuities, or freezing their last submitted density, which ignores new information, our approach updates each forecaster's latent predictive state through the implied conditional structure of the panel. This isolates genuine performance differences from participation effects and ensures that the aggregate density evolves smoothly in response to economic news rather than administrative turnover. The Bayesian coherence requirement yields explicit, local entry and exit adjustment rules that preserve the learning dynamics of synthesis weights and avoid artificial discontinuities in predictive means and variances.

We evaluate the proposed approach in real-time forecasting experiments using nearly 25 years of inflation density forecasts. Relative to equal-weight pooling-- a benchmark that is often difficult to outperform in practice-- the coherent synthesis approach improves point forecast accuracy and delivers substantially better-calibrated density forecasts. These gains are most pronounced during periods of high volatility and elevated forecaster turnover. From an economic perspective, the improved calibration implies that movements in forecast dispersion and tail risk more closely reflect genuine shifts in uncertainty rather than changes in panel composition, a distinction that is particularly important for central banks' real-time assessment of inflation risks.

The paper makes three contributions. First, it formalizes forecast combination under forecaster turnover and missingness for predictive distributions, and shows empirically that common ad hoc fixes can materially distort combined uncertainty and confound evaluation in a major institutional survey. Second, it develops coherence-based entry and exit adjustments within dynamic Bayesian predictive synthesis, yielding a principled and operational aggregation rule under arbitrary participation patterns. Third, it provides an application-driven empirical assessment showing that coherence-based adjustments improve predictive accuracy and enhance the economic interpretability of uncertainty and tail-risk dynamics in real time.

The remainder of the paper is organized as follows. Section~2 introduces the SPF setting and the practical pathologies that arise under standard pooling when participation is sporadic. Section~3 states the coherent probabilistic formulation that motivates our approach. Section~4 develops the coherence-based entry and exit operators within dynamic BPS. Section~5 presents the forecasting design, benchmarks, and empirical results. Section~6 discusses practical implications and extensions, and Section~7 concludes. Supplementary Material includes computational details and robustness analyses regarding specification, forecaster set size, and dataset (unemployment rate).

\subsection*{Related Literature}

The forecast combination literature, dating to \citet{BatesGranger1969}, has developed linear pooling, shrinkage weights, BMA, and optimal combination under loss-based criteria \citep[e.g.,][]{Clemen1989,Timmermann2004,jore2010combining,Geweke2011}. For density forecasting specifically, several approaches combine forecast densities \citep[e.g.][]{HallMitchell2007,Amisano2007,Geweke2011,Billio2012,Aastveit2014,Pettenuzzo2015,Negro2016,aastveit2018evolution}. These methods typically treat forecasters as fixed and fully observed, offering little guidance when the contributor set changes over time.

Bayesian predictive synthesis (BPS) provides a coherent framework for combining forecast densities by introducing latent agent-specific predictive states and a synthesis function \citep{mcalinn2019dynamic}. BPS has been applied to multivariate macro forecasting \citep{mcalinn2017multivariate}, mixed-frequency nowcasting \citep{mcalinn2021mixed}, outcome-dependent pooling \citep{JohnsonWest2022}, and decision synthesis for monetary policy \citep{tallman2024bayesian,ChernisEtAl2024MonPolDS,adrian2025scenario}. In survey environments where the set of available forecasts changes unpredictably, BPS offers a natural route to maintain probabilistic coherence.

A small body of work studies forecast panels with missing data \citep[e.g.,][]{Genre2013,dovern2015multivariate}, but focuses on modeling individual forecasts or imputing missing responses rather than on coherent combination. Standard treatments rely on ad hoc imputations, EM-style smoothing, or listwise deletion-- none of which preserves the conditional structure implied by a coherent Bayesian rule when the observed set varies over time. The institutional forecasting literature has focused largely on assessing biases or disagreement in surveys \citep[e.g.,][]{clements2010explanations,dovern2011accuracy,boero2015measurement}, rather than on coherent combination under sporadic participation.

%
%
\section{Forecasting with a Professional Forecaster Survey} \label{sec:application}
\begin{figure}[!t]
    \centering \setlength{\tabcolsep}{0\textwidth}
 \begin{tabular}{C{.5\textwidth}C{.5\textwidth}}   
    \includegraphics[width=0.49\textwidth]{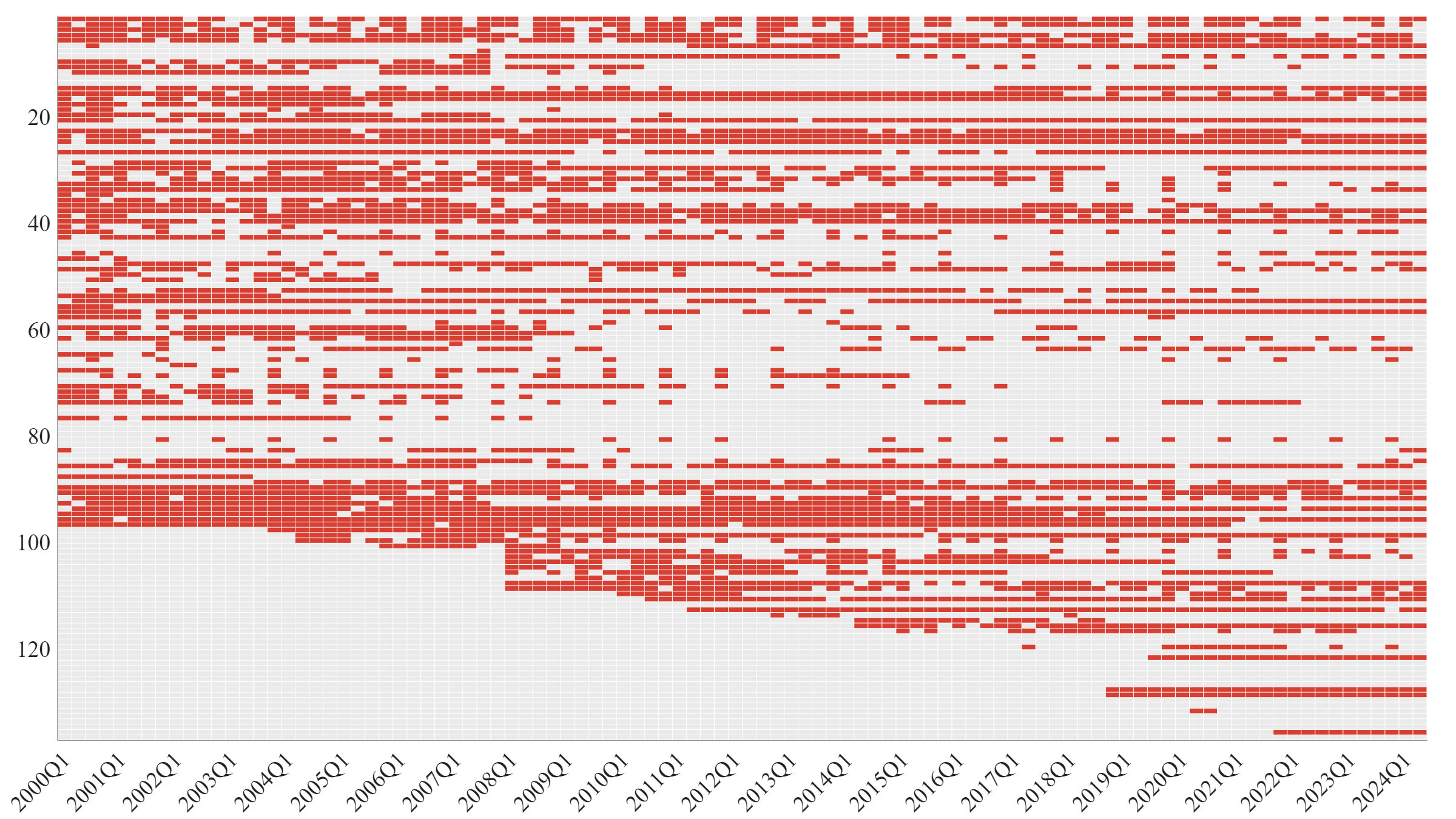}&    
    \includegraphics[width=0.49\textwidth]{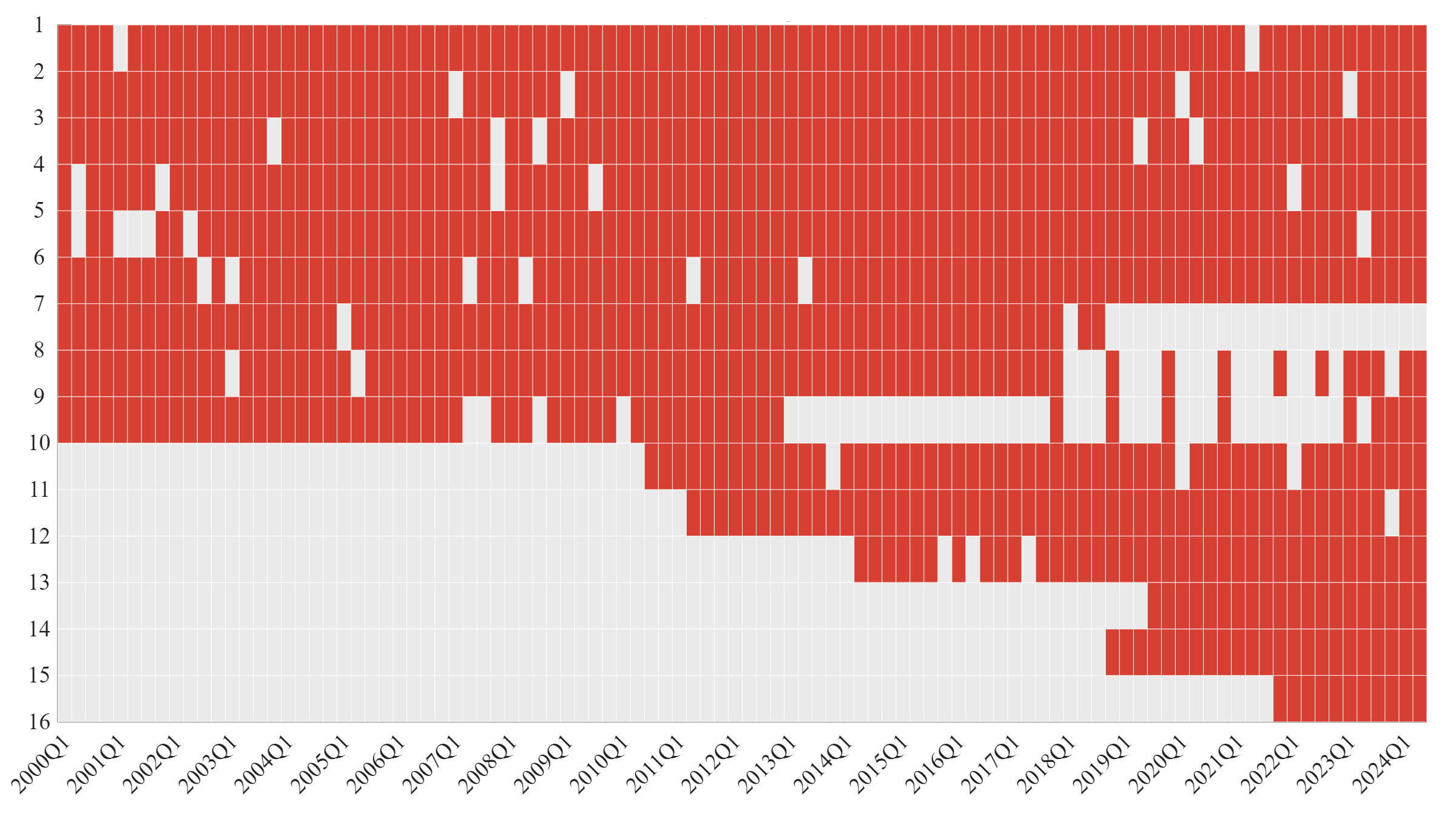}\\
    \small All forecasters &  \small Forecasters used in application \\
\end{tabular}
    \caption{Participation of expert forecasters over time. Left panel: participation of the full panel. Right panel: participation of the subset panel used for analysis.
    Light gray indicates inactivity and red denotes active contributions.}
    \label{fig:participation}
\end{figure}

This paper aims to address a practical problem that arises in institutional forecasting systems that rely on professional forecaster surveys: how to construct a single, real-time predictive distribution when the set of contributing forecasters is unstable. In our application, the objective is to produce quarterly inflation density forecasts by combining individual predictive distributions from the European Central Bank's Survey of Professional Forecasters (SPF), a widely used input for inflation monitoring and communication. A defining feature of this environment is that participation is not balanced, as forecasters enter and exit over time, skip rounds, and reappear after long gaps, so that the panel of available density forecasts changes from quarter to quarter (Figure~\ref{fig:participation}).\ \ \textit{This is the application problem.} Standard pooling rules typically presume a fixed contributor set, and when applied mechanically to a changing panel, they can generate artificial movements in combined predictive moments driven by changes in the conditioning set rather than new information.

\subsection{Why turnover and missingness are first-order features}\label{subsec:firstorderfeature}
Figure~\ref{fig:participation} provides a heatmap of forecaster activity over time. 
Since the inception of the survey, more than 130 distinct forecasters have participated, producing an extremely sparse and irregular panel (left). For real-time evaluation, we focus on a pre-specified \emph{core panel} that provides enough history to initialize and score forecasting rules while still retaining substantial entry/exit and intermittent participation (right).  Concretely, we select the forecasters with the largest total number of non-missing density submissions over the sample, subject to having sufficient observations in the initial training window for initialization.  This selection yields a core panel of $J=16$ forecasters that is information-rich enough to support stable real-time benchmarking, while still exhibiting long participation gaps and substantial time-varying turnover (Figure~\ref{fig:participation}, right chart).
Section~\ref{app:robust_20} of the Supplementary Material verifies that the main findings are robust to expanding the panel to $J=20$ forecasters.

Three features of Figure~\ref{fig:participation} are particularly relevant for real-time aggregation. First, entry and exit are frequent: the active set $A_t$ is not slowly varying, and there is no stable or steady-state panel. Second, missingness is persistent: many forecasters are absent for extended stretches rather than missing a single round. Third, participation varies over time in ways plausibly related to the macroeconomic environment, so panel instability is often most pronounced precisely when forecast uncertainty is most consequential.

\begin{figure}[!t]
    \centering
    \begin{subfigure}{0.45\linewidth}
        \centering
        \includegraphics[width=\linewidth]{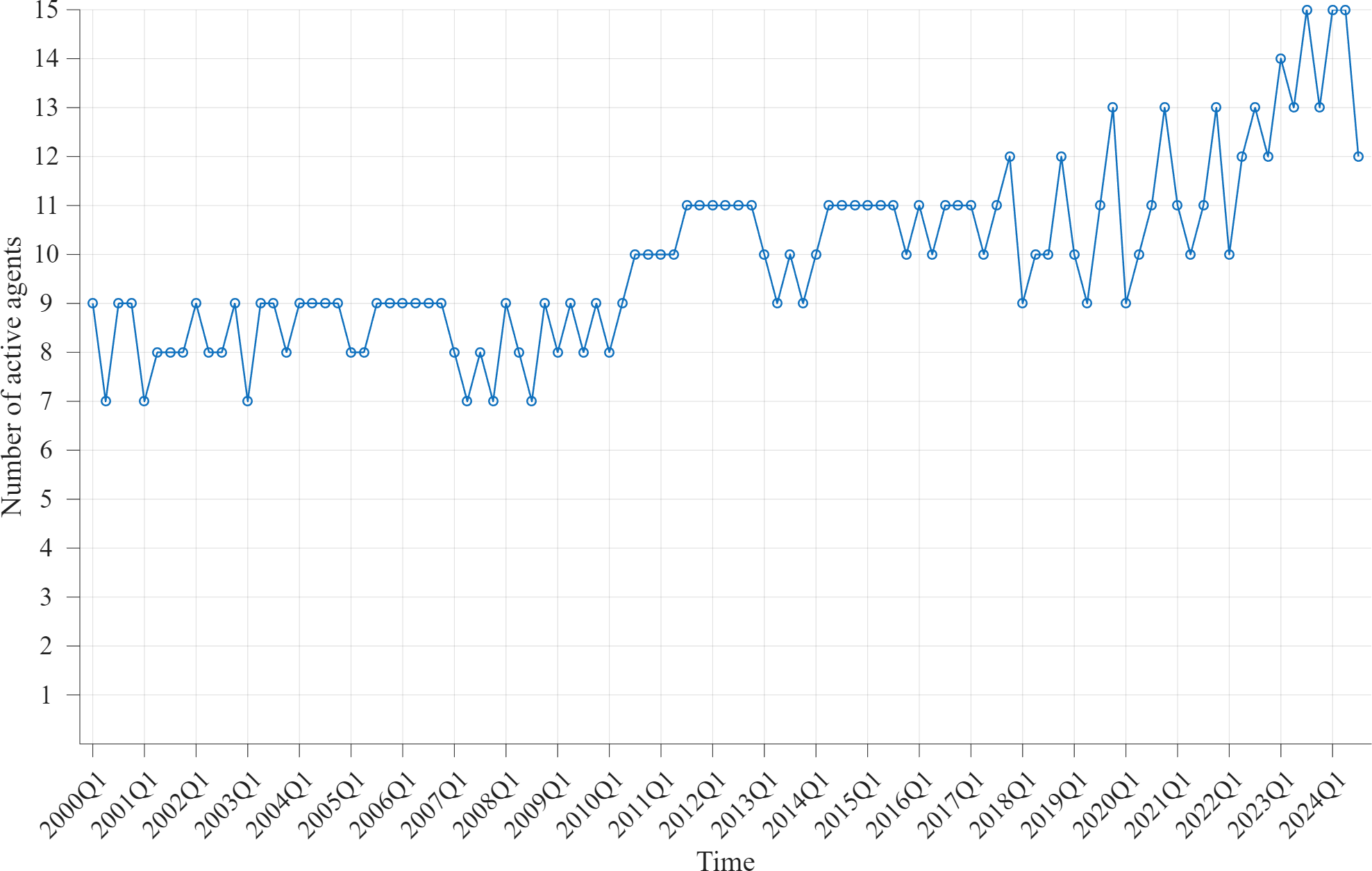}
        \caption{Number of active experts over time.}
    \end{subfigure}
    \begin{subfigure}{0.45\linewidth}
        \centering
        \includegraphics[width=\linewidth]{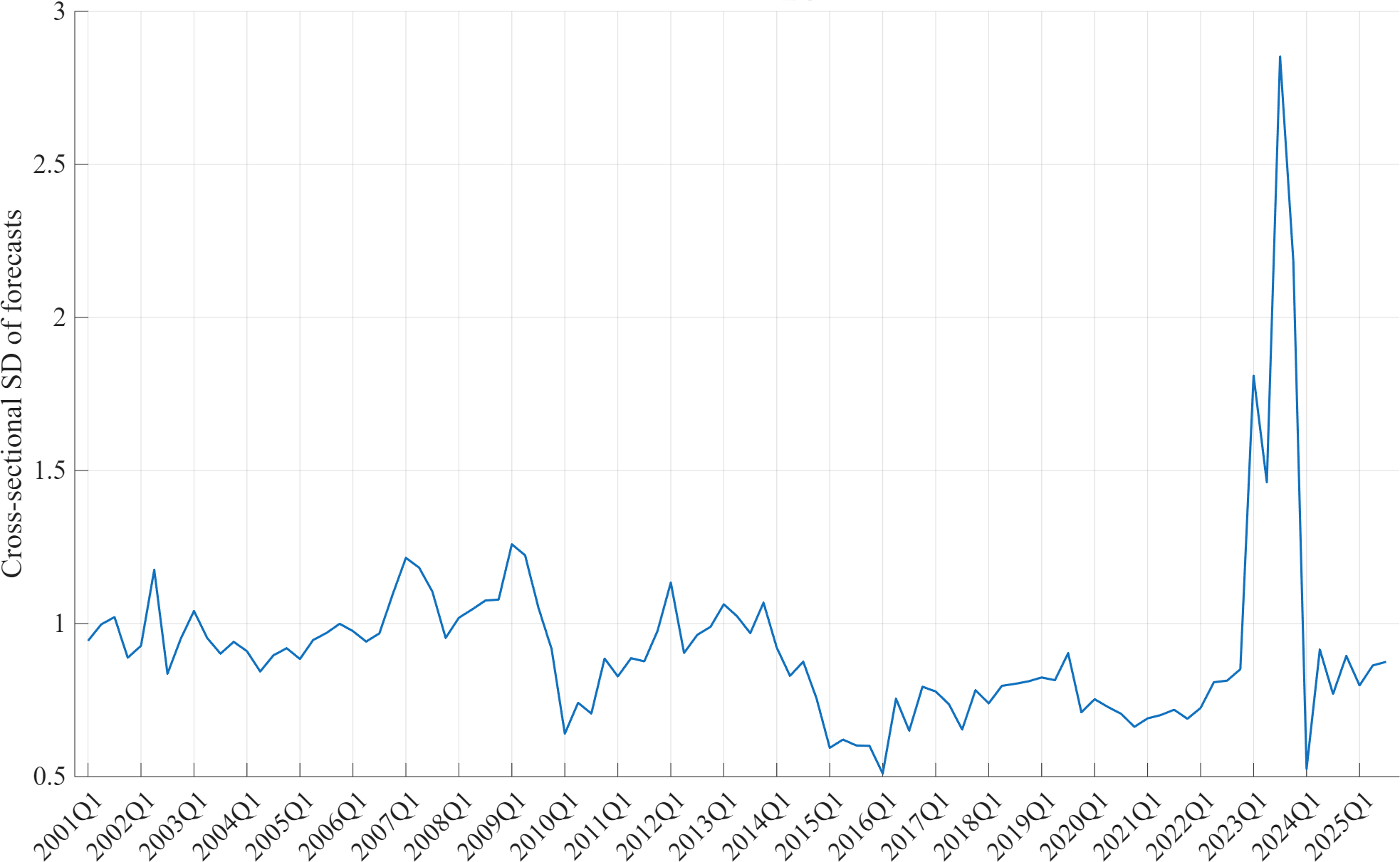}
        \caption{Forecast dispersion among active experts.}
    \end{subfigure}

    \vspace{0.5em}

    \begin{subfigure}{0.45\linewidth}
        \centering
        \includegraphics[width=\linewidth]{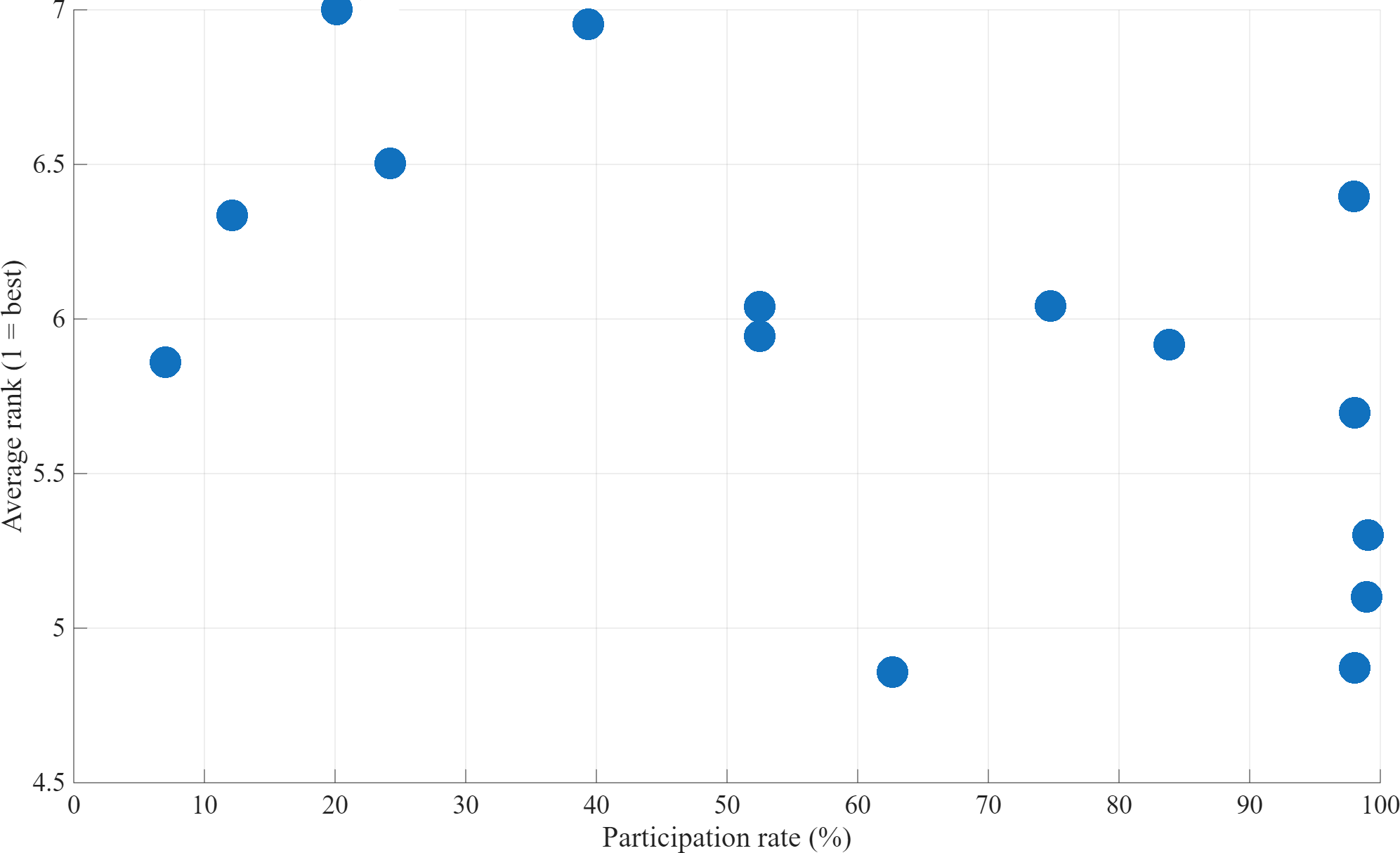}
        \caption{Forecast error on participation rate.} 
    \end{subfigure}

    \caption{Descriptive characteristics of the central-bank forecasting panel. }
    \label{fig:descriptives}
\end{figure}

\begin{figure}[!t]
    \centering
    \includegraphics[width=0.5\linewidth]{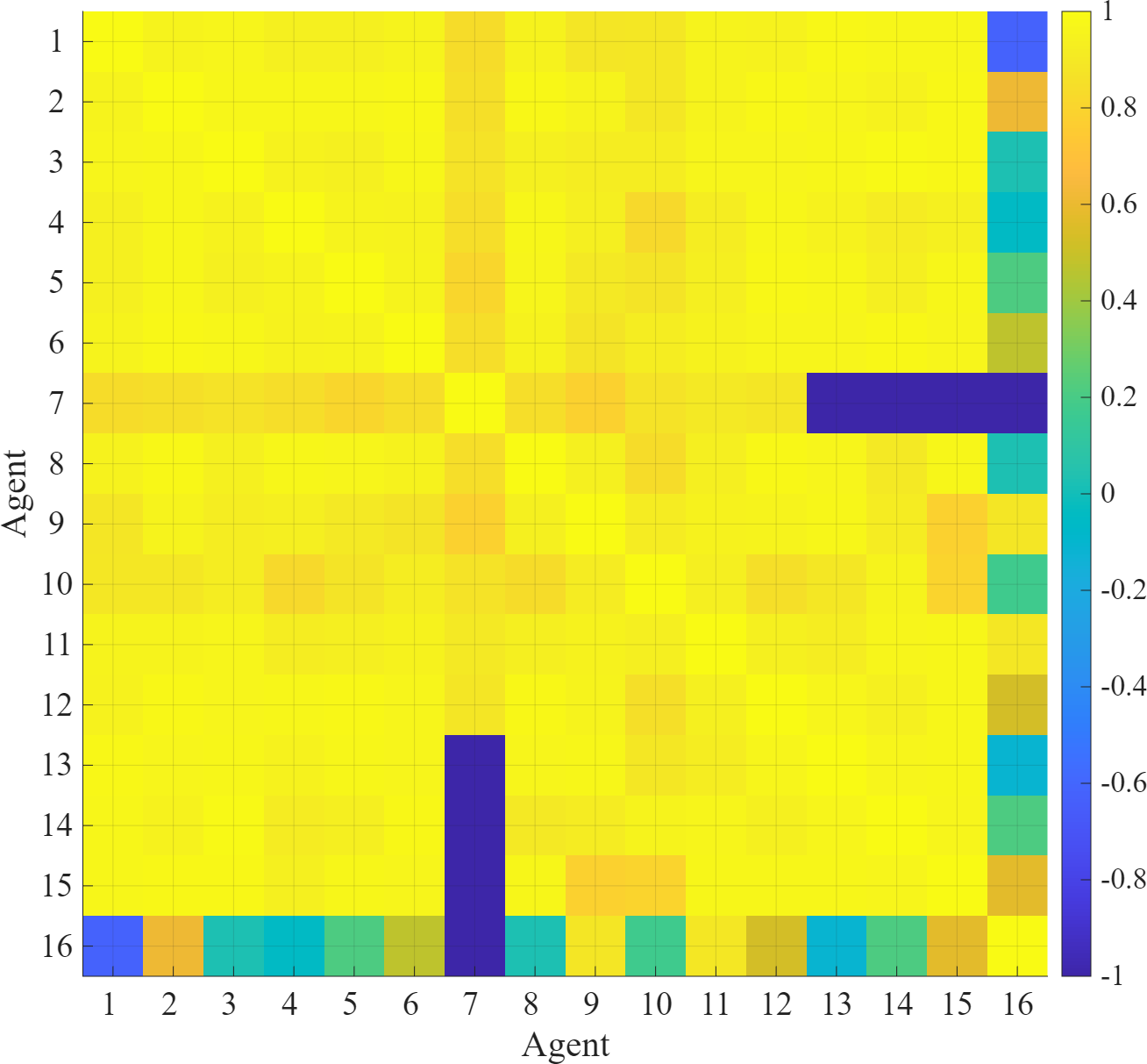}
    \caption{Correlation matrix of experts' forecast errors. }
    \label{fig:corr}
\end{figure}

These features are not innocuous bookkeeping issues. Figure~\ref{fig:descriptives} shows substantial time variation in the number of active experts (Panel (a)), pronounced dispersion among participating forecasts (Panel (b)), and a negative relationship between participation rate (share of quarters active) and predictive error (Panels  (c)), suggesting that participation is not random. Figure~\ref{fig:corr} further documents strong dependence in forecast errors across experts, reinforcing that forecast diversity coexists with correlation and that aggregation methods must account for both skill heterogeneity and cross-sectional dependence. This is also something we can leverage in our analysis, since if two forecasters are highly correlated and one drops out, it is natural to think that the change in weights should reflect that high correlation.

\subsection{How institutions typically combine survey forecasts}
In practice, the dominant operational baseline in survey-based forecasting is to average whatever submissions are available prior to time $t$: the equal-weight (EW) combination rule. In our empirical analysis, EW serves as the main reference benchmark, and it is well known to be difficult to outperform in many forecasting contexts \citep[e.g.,][]{Genre2013}. 

Related approaches depart from EW by estimating combination weights via frequentist scoring rules under stabilizing constraints (e.g., nonnegativity and simplex constraints), for both point and density forecasts; see \citet{ConflittiDeMolGiannone2015OptimalCombination} and the regularized ``partially-egalitarian'' LASSO approach of \citet{diebold2019machine}. A second class of pragmatic fixes imputes missing submissions-- such as last-observation-carried-forward or forecaster-specific mean imputation-- prior to averaging. A third class estimates weights from historical forecast errors, for example using inverse mean-squared-error weights computed over rolling windows. 

Despite their differences, these strategies implicitly treat the set of forecasters as effectively fixed over the estimation window and are therefore sensitive to instability in $A_t$. Our focus is on the complementary (and empirically central) issue that in many surveys the forecaster set itself evolves rapidly through entry, exit, and intermittent participation, so that aggregation rules must be coherent to the changing information set rather than to an artificially stabilized panel.

\subsection{The application problem: why naive fixes distort density forecasts}
Turnover and missingness create a mismatch between the institutional objective and what naive pooling delivers. The institutional target is a single quarterly predictive density whose movements reflect new information about inflation and changes in forecasters' beliefs, not administrative changes in who happened to respond. When the active set $A_t$ changes, however, common operational fixes implicitly change the \emph{conditioning set} underlying the pooled object.

Two mechanisms are particularly problematic for density forecasting. First, imputing missing densities introduces synthetic information and effectively modifies the joint predictive structure underpinning the pooled density, especially when absences are long or permanent. Second, dropping missing forecasters and renormalizing weights reallocates influence across remaining forecasters and can produce artificial jumps in combined predictive means and variances even when the underlying forecast environment is stable. In a density-forecasting context, these artifacts are consequential: they alter perceived risk levels, distort tail probabilities, and confound backtesting of calibration and sharpness because forecast changes may be driven by participation rather than learning.

\subsection{Requirements for survey-based density combination}
The SPF setting suggests concrete requirements for a usable combination procedure: (i) the combined density should update by conditioning on observed forecasts rather than by mechanically renormalizing weights or inserting ad hoc imputations; (ii) pooled dispersion and tail-risk measures should not jump mechanically at entry and exit times; (iii) the method should not require specifying a structural model for participation; and (iv) the procedure should support meaningful inference about forecaster value added even when participation is intermittent. These requirements are motivated directly by the empirical features of the panel: frequent entry and exit, persistent missingness, and time-varying participation.

%
%
\section{Problem Formulation} \label{sec:problemform}
We now translate the practical requirements of Section~\ref{sec:application} into a probabilistic framework, formalizing what constitutes a coherent update of a combined predictive distribution when only a subset of forecasters submits densities and the active set varies over time.

\subsection{Observed forecast densities and a time-varying active set}
Let $y_t$ denote target year-ahead inflation measured at quarter $t$, and let $h_{j,t}(\cdot)$ denote the density forecast for $y_{t}$ submitted by forecaster $j$ before time $t$. There is a pool of up to $J$ potential forecasters. Before time $t$, forecaster $j$ may provide a full predictive density for $y_{t}$,
\[
h_{j,t}(x) \;\equiv\; p_j\!\left(y_{t}=x_t | \mathcal{I}_{j,t}\right),
\]
where $\mathcal{I}_{j,t}$ denotes forecaster $j$'s information set regarding time $t$. Let $A_t \subseteq \{1,\ldots,J\}$ denote the set of active forecasters at time $t$; those who submit a density forecast. 
Throughout, double subscripts of the form $H_{A_t,t}$ denote the restriction of quantity $H$ to the forecaster subset $A_t$ at time $t$.
We observe the collection
\[
H_{A_t,t} \;\equiv\; \{h_{j,t}(\cdot): j\in A_t\},
\]
while densities for $j\notin A_t$ are unobserved. 

A decision maker (DM) seeks a single predictive distribution for $y_{t}$ given past outcomes and the densities observed before time $t$ (say, e.g., $t-1$). The key complication is that $A_t$ varies over time, so a coherent model must specify how the DM’s predictive distribution updates when only $H_{A_t,t}$ is observed.

\subsection{Predictive synthesis with latent predictive states}
Bayesian predictive synthesis (BPS) represents each forecaster's density as information about a latent predictive state $x_{j,t}$ and combines these states through a synthesis function. Let $\x_{t}=(x_{1,t},\ldots,x_{J,t})'$ collect the latent states, and let $m_{t}(\x_{t})$ denote the DM's joint predictive prior for $\x_t$ (prior to observing the forecasters' densities before time $t$). Let $p_t(y_t)$ denote the DM's predictive prior for $y_{t}$ based on information available up to the time to make predictions for time $t$. The DM specifies a synthesis function $\alpha_t(y_t | \x_t,\Phi_t)$, with parameters $\Phi_t$, that maps latent states into a predictive density for $y_{t}$.
Under a fully balanced panel-- when all $J$ densities are observed-- the BPS posterior predictive takes the form
\begin{equation}
p(y_{t}| H_t, y_{1:t-1}) \;=\; \int \alpha_t(y_{t}| \x_t,\Phi_t)\,
\prod_{j=1}^J h_{j,t}(x_{j,t})\, dx_t,
\label{eq:bps_full}
\end{equation}
where $H_t=\{h_{1,t},\ldots,h_{J,t}\}$.

\subsection{The coherence (consistency) condition}
The formulation in \eqref{eq:bps_full} is Bayesian-coherent only if the DM's prior predictive distribution $p_t(y_t)$ is consistent with the joint prior $m_t(\x_t)$ through the synthesis function:
\begin{equation}
p_t(y_t) \;=\; \int \alpha_t(y_t| \x_t,\Phi_t)\, m_t(\x_t)\, d\x_t .
\label{eq:consistency}
\end{equation}
Equation \eqref{eq:consistency} is the coherence (or consistency) condition: the DM's prior for $y_{t}$ must equal the marginal distribution implied by the joint model for $(y_{t},\x_t)$. This requirement is typically innocuous under a fixed and fully observed panel, where priors can be chosen to satisfy it directly. Under turnover and missingness, however, coherence becomes decisive because the condition must continue to hold when only a subset of forecast densities is observed and the active set $A_t$ changes over time.

\subsection{Partial observation: coherent updating with only $H_{A_t,t}$}
When only $H_{A_t,t}$ is observed, BPS treats missing forecasters as unobserved components of the latent predictive state rather than as contributors that disappear from the model. Write $\x_t=(x_{A_t,t},x_{A_t^c,t})$ for the partition of latent states into active and inactive components. Coherent updating replaces the prior for the observed components with the reported densities while leaving the inactive components governed by their predictive prior.

Define the restricted synthesis function obtained by marginalizing over inactive states using the DM's conditional prior distribution:
\begin{equation}
\alpha_{A_t,t}(y_t| x_{A_t,t},\Phi_t)
\;=\;
\int \alpha_t\!\left(y_t| x_{A_t,t},x_{A_t^c,t},\Phi_t\right)\,
m_t(x_{A_t^c,t}| x_{A_t,t})\, dx_{A_t^c,t}.
\label{eq:alpha_restricted}
\end{equation}
The coherent posterior predictive distribution given only the active-set densities is then
\begin{equation}
p(y_{t}| H_{A_t,t}, y_{1:t-1})
\;=\;
\int \alpha_{A_t,t}(y_{t}| x_{A_t,t},\Phi_t)\,
\prod_{j\in A_t} h_{j,t}(x_{j,t})\, dx_{A_t,t}.
\label{eq:bps_partial}
\end{equation}

Equations \eqref{eq:alpha_restricted}--\eqref{eq:bps_partial} formalize how a missing forecaster should be handled in a coherent probabilistic framework: as marginalization over an unobserved latent state rather than as a change in model dimension or a renormalization of weights. This is the probabilistic object exploited in the next section to derive explicit entry and exit operators in a dynamic BPS model, ensuring that posterior learning and predictive uncertainty evolve smoothly as $A_t$ changes over time.

%
%
\section{Coherent Updating Under Entry and Exit} \label{sec:entry_exit}
This section specializes the coherence principle from Section~\ref{sec:problemform} to the dynamic BPS model of \citet{mcalinn2019dynamic} and derives explicit entry--exit operators. The key idea is that the latent predictive state and synthesis coefficients retain fixed dimension $J$ across time, while entry and exit only change which components of the latent state are observed (and hence enter the likelihood) before time $t$.

\subsection{Joint Gaussian updating with time-varying expert sets}
Let $\btheta_t=(\theta_{0,t},\theta_{1,t},\ldots,\theta_{J,t})'$ denote synthesis coefficients at time $t$, and let $\x_t=(x_{1,t},\ldots,x_{J,t})'$ denote latent expert-specific predictive states. Under the dynamic BPS model of \citet{mcalinn2019dynamic}, conditional on $(\btheta_t,\x_t)$ the synthesis function is Gaussian and linear:
\begin{equation}
    y_t \;=\; \theta_{0,t} + \theta_{1:J,t}' \x_t + \nu_t,
\qquad \nu_t\sim N(0,v_t).
\label{eq:dlm_meas}
\end{equation}
Given past data and past forecast information, the pair $(\btheta_t,\x_t)$ has a predictive prior (Gaussian under the state evolution and the Gaussian forecast-state specification). Entry and exit are handled by ensuring that when only a subset of experts is observed, the predictive distribution for $y_t$ is obtained by coherent Bayesian conditioning on that subset-- without changing the dimension of $(\btheta_t,\x_t)$.

Let $A_t\subseteq\{1,\ldots,J\}$ denote the active expert set at time $t$, and let $A_t^c$ denote its complement. Write $\x_t=(x_{A_t,t},x_{A_t^c,t})$ and similarly $\btheta_t=(\theta_{A_t,t},\theta_{A_t^c,t})$. The observed forecast densities before time $t$ are $\{h_{j,t}:j\in A_t\}$; inactive experts correspond to unobserved components of the predictive state, not parameters that disappear from the model.

\subsection{Exit as marginalization}
Suppose some experts are active at $t-1$ but not at $t$, i.e.\ $j\in A_{t-1}\setminus A_t$. Since the time-$t$ likelihood only depends on $x_{A_t,t}$ (through the observed forecast information) and on the active coefficients, coherent updating treats $(x_{A_t^c,t},\theta_{A_t^c,t})$ as latent components governed by their predictive prior. Equivalently, the synthesis function is updated by marginalizing over the inactive states under the decision maker's prior conditional distribution:
\begin{align}
p_t(y_t | H_{A_t,t})
&=\int \alpha_{A_t,t}(y_t| x_{A_t,t})\prod_{j\in A_t} h_{j,t}(x_{j,t})\,dx_{A_t,t},
\label{eq:exit_post_pred}\\
\alpha_{A_t,t}(y_t| x_{A_t,t})
&=\int \alpha_t(y_t| x_{A_t,t},x_{A_t^c,t})\, m_t(x_{A_t^c,t}| x_{A_t,t})\,dx_{A_t^c,t}.
\label{eq:exit_alpha_marg}
\end{align}
This is exactly the ``drop an expert'' operation implied by coherence: it is Bayesian marginalization of an unobserved latent state, not a change in model dimension and not a renormalization rule.

\paragraph{Closed-form Gaussian consequences.}
Assume (i) the active-set forecast states are Gaussian with
\[
x_{A_t,t}| H_{A_t,t}\sim N(h_{A_t,t},H_{A_t,t}),
\]
(with $H_{A_t,t}$ diagonal under independent forecast-state sampling across active experts), and (ii) the predictive prior for the full latent state is jointly Gaussian:
\[
\begin{pmatrix}x_{A_t,t}\\ x_{A_t^c,t}\end{pmatrix}
\sim N\!\left(
\begin{pmatrix}m_{A_t,t}\\ m_{A_t^c,t}\end{pmatrix},
\begin{pmatrix}
M_{A_t,t} & M_{A_t,A_t^c,t}\\
M_{A_t^c,A_t,t} & M_{A_t^c,t}
\end{pmatrix}
\right).
\]
Let
\[
B_t \;=\; M_{A_t,A_t^c,t}\,M_{A_t,t}^{-1}.
\]
Then marginalizing the inactive states induces an adjusted linear-Gaussian synthesis for $y_t| H_{A_t,t}$ with
\begin{align}
\mathbb{E}[y_t| H_{A_t,t}]
&= \tilde\theta_{0,t} + \tilde\theta_{A_t,t}'\,h_{A_t,t},
\label{eq:exit_mean}\\
\mathrm{Var}(y_t| H_{A_t,t})
&= v_t
 + \theta_{A_t,t}'H_{A_t,t}\theta_{A_t,t}
 + \theta_{A_t^c,t}'M_{A_t^c,t}\theta_{A_t^c,t}
 + \theta_{A_t^c,t}'\!\left(B_t'(H_{A_t,t}-M_{A_t,t})B_t\right)\theta_{A_t^c,t},
\label{eq:exit_var}
\end{align}
where the effective (coherence-implied) coefficient adjustments are
\begin{align}
\tilde\theta_{0,t}
&=\theta_{0,t}+\theta_{A_t^c,t}'\!\left(m_{A_t^c,t}-B_t'h_{A_t,t}\right),
\label{eq:exit_theta0}\\
\tilde\theta_{A_t,t}
&=\theta_{A_t,t}+B_t\,\theta_{A_t^c,t}.
\label{eq:exit_thetaA}
\end{align}
Equations \eqref{eq:exit_theta0}--\eqref{eq:exit_thetaA} make the mechanism transparent: exit reallocates the departing experts' influence through the prior cross-covariances (via $B_t$), and shifts the intercept according to how the realized active-set mean $h_{A_t,t}$ differs from the prior-implied mean for the inactive states.

The intuition is as follows. When a forecaster exits, their influence on the combined forecast does not simply vanish; instead, it is redistributed to the remaining forecasters in proportion to how well their latent states predict the exiting forecaster's state. The matrix $B_t$ captures exactly this predictive relationship through the prior cross-covariances. If an exiting forecaster is highly correlated with a continuing forecaster, most of the exiting forecaster's synthesis weight transfers to that continuing forecaster (via equation \eqref{eq:exit_thetaA}). If the exiting forecaster is largely idiosyncratic, their weight is absorbed into the intercept (via equation \eqref{eq:exit_theta0}). This structure ensures that exit does not create artificial discontinuities: the combined forecast changes smoothly because the information previously contributed by the exiting forecaster is now inferred-- at appropriately greater uncertainty-- from the remaining panel.

\paragraph{Implementation (transport view).}
In the MCMC/filtering implementation, exit is realized by drawing from the full predictive Gaussian for $\btheta_t$ and
then applying a linear (Gaussian) transport map so that the implied posterior moments match the coherent marginal
distribution after removing the exiting experts from the observed likelihood before time $t$. The exiting experts remain in
the state vector; coherence is enforced by marginalization and moment-preserving adjustment in the active subspace.

\subsection{Entry as conditional projection}
Now suppose experts enter before time $t$, so $j\in A_t\setminus A_{t-1}$. Entry is the reverse problem: the enlarged
model for time $t$ must reduce to the previous predictive distribution when the entering experts are marginalized out.
Formally, letting $A_{t-1}$ denote the continuing experts and $A_{t-1}^c$ the entering experts (so that $A_t=A_{t-1}\cup A_{t-1}^c$),
coherence requires that the pre-entry predictive be recovered by marginalization:
\begin{align}
p_t(y_t)
&=\int \alpha_{A_{t-1},t}(y_t| x_{A_{t-1},t})\, m_t(x_{A_{t-1},t})\,dx_{A_{t-1},t}
\label{eq:entry_consist_1}\\
&=\int\!\!\int \alpha_t(y_t| x_{A_{t-1},t},x_{A_{t-1}^c,t})\,
m_t(x_{A_{t-1},t},x_{A_{t-1}^c,t})\,dx_{A_{t-1},t}\,dx_{A_{t-1}^c,t}.
\label{eq:entry_consist_2}
\end{align}
Unlike exit, entry is not uniquely identified from \eqref{eq:entry_consist_1}--\eqref{eq:entry_consist_2} alone:
adding new coefficients and states introduces degrees of freedom. We therefore adopt a pragmatic, coherence-preserving
construction: place a prior on the entering coefficients and extend the synthesis function by conditional projection.

Assume the same Gaussian setup as above with partition $(A_{t-1},A_{t-1}^c)$ and define
\[
B_t \;=\; M_{A_{t-1},A_{t-1}^c,t}\,M_{A_{t-1},t}^{-1},
\qquad
M_{A_{t-1}^c| A_{t-1},t}
\;=\;
M_{A_{t-1}^c,t}-M_{A_{t-1}^c,A_{t-1},t}M_{A_{t-1},t}^{-1}M_{A_{t-1},A_{t-1}^c,t}.
\]
Let $(\tilde\theta_{0,t},\tilde\theta_{A_{t-1},t},\tilde v_t)$ denote the \emph{known} parameters of the pre-entry synthesis
(i.e., the model implied by the continuing experts). Given a sampled value of the entering coefficients
$\theta_{A_{t-1}^c,t}$ (from its prior), coherence is enforced by solving for the remaining parameters so that
marginalizing out the entering states recovers the pre-entry predictive:
\begin{align}
\theta_{0,t}
&= \tilde\theta_{0,t} - \theta_{A_{t-1}^c,t}'\!\left(m_{A_{t-1}^c,t}-B_t' m_{A_{t-1},t}\right),
\label{eq:entry_theta0}\\
\theta_{A_{t-1},t}
&= \tilde\theta_{A_{t-1},t} - B_t\,\theta_{A_{t-1}^c,t},
\label{eq:entry_thetaA}\\
v_t
&= \tilde v_t - \theta_{A_{t-1}^c,t}'\,M_{A_{t-1}^c| A_{t-1},t}\,\theta_{A_{t-1}^c,t}.
\label{eq:entry_v}
\end{align}
In words: entry activates new coefficients while adjusting the intercept, the continuing coefficients, and the residual
variance so that the enlarged synthesis function is a coherent extension of the pre-entry model.

\paragraph{Implementation.}
In the sampler, entry is implemented by (i) specifying a prior mean/variance for the entering coefficients
$\theta_{A_{t-1}^c,t}$ (optionally informed by empirical hyperparameters), (ii) drawing from the full predictive Gaussian,
and (iii) applying a linear transformation that enforces the conditional mean and covariance implied by
\eqref{eq:entry_theta0}--\eqref{eq:entry_v}. The result is smooth activation of the new coefficient and continuity of the
predictive distribution across entry events.

\subsection{Summary}
Exit corresponds to coherent marginalization: inactive experts remain in the state vector and evolve under their predictive prior, while the synthesis function is updated by integrating out their unobserved states. Entry corresponds to coherent conditional projection: new experts are added so that marginalizing them out recovers the pre-entry predictive distribution. The explicit operators are finite-sample realizations of these transformations via Gaussian draws followed by moment-preserving linear maps. As a result, the posterior for $(\btheta_t,\x_t)$ retains fixed dimension and the predictive distribution evolves smoothly even under substantial panel turnover.

%
%
\section{Empirical Results: Forecasting with a Central-Bank Expert Panel}
We evaluate the proposed coherent Bayesian synthesis model using inflation density forecasts from the European Central Bank’s Survey of Professional Forecasters. As discussed in Section~\ref{sec:application}, there is substantial turnover among survey participants, with only a handful of experts appearing in more than half the sample, several contributing only intermittently, and many disappearing entirely for long stretches. This empirical setting provides a direct test of whether coherence-based updating improves predictive performance and stability relative to standard pooling and imputation-based benchmarks.
To demonstrate robustness, in Supplementary Material~\ref{app:unemp}, we mirror the analysis presented in this section using unemployment rate density forecasts from the same SPF survey.

\subsection{Data structure and experimental setup}\label{subsec:exp_setup}
The dataset contains $T=99$ quarters of year-ahead inflation forecasts from $J=16$ forecasters (we find the results to be consistent when we vary the number of forecasters, see Supplementary Material~\ref{app:robust_20}). At each quarter $t$, only a subset $A_t \subseteq \{1,\ldots,J\}$ of forecasters submit density forecasts. The panel is treated exactly as observed: if forecaster $j$ does not report at time $t$, then $j\notin A_t$ and no forecast from that source is available at that date.

At each quarter, we condition on the information set $D_t=\{y_{1:t-1}, \x_{1:t}\}$, where $y_t$ denotes realized inflation and $\x_t$ collects the latent expert signals implied by the currently active panel. Models are estimated recursively, using only information available at the time forecasts are issued, and the predictive distribution $p(y_{t+4}| D_t)$ is generated for $t=1,\ldots,T-1$.

For initialization, we use the first 42 quarters (2000:Q1–2010:Q2). During this initial period only, missing expert forecasts are linearly interpolated to provide common starting values across all methods. This device ensures that every method-- including the benchmarks-- begins the evaluation period from an identical information state, so that cumulative performance metrics start at zero and subsequent differences reflect only out-of-sample behavior. Importantly, this interpolation is applied identically to all methods and affects only the pre-evaluation training window; no imputation of any kind is used during the 57-quarter evaluation period, where entry and exit are handled exactly as observed. This design prevents imputation artifacts in the training set from propagating into the evaluation metrics in ways that would be difficult to diagnose or attribute.

\subsection{Benchmark methods} \label{subsec:bench_methods}
The empirical objective is to assess whether a coherent treatment of forecaster entry, exit, and intermittent participation improves forecasting performance relative to commonly used aggregation schemes. We compare the proposed model with the following benchmark methods:
\begin{enumerate}
  \item \textbf{Equal-weight (EW) combination.} At each $t$ the forecast is the simple average of all currently available expert forecasts, $y_{t| t-1}^{\text{EW}} = |\mathcal{A}_t|^{-1} \sum_{j\in{A}_t} f_{t| t-1}^{(j)}$.  This method is widely used in institutional settings and serves as the benchmark  model for our forecast evaluation. {EW is difficult to outperform in practice and therefore provides a demanding benchmark.}
  \item \textbf{Imputation-based equal weights.} We also consider equal-weight combinations applied to panels in which missing forecasts are first imputed using simple rules, including last-observation-carried-forward (LOCF) and agent-specific mean imputation (ASMI). These approaches construct a balanced panel prior to aggregation but do not enforce coherence of the resulting predictive distribution.
  \item \textbf{Inverse-MSE.} As an example of performance-based weighting, we consider linear pools with weights proportional to the inverse of rolling-window mean squared forecast errors. Specifically, weights satisfy $w_{t,j} \propto \widehat{\mathrm{MSE}}^{-1}_{t,j}$. This approach treats the set of experts as fixed within the estimation window and is therefore sensitive to panel instability.
\end{enumerate}

A natural question is why standard BPS-- without the entry and exit operators-- is not included as a benchmark. The answer is that standard BPS, like most advanced forecast combination methods, requires a fixed panel of forecasters and cannot be applied when the contributor set varies over time. Because no forecaster in our sample participates in every quarter, running standard BPS would require substantial imputation throughout the evaluation period, confounding any comparison. The benchmarks above represent the operational alternatives actually available when participation is sporadic: pool whatever is observed, impute and then pool, or estimate weights from available history.

\subsection{Estimation details for BPS}
The coherent BPS model is estimated using the MCMC algorithm described in Supplementary Material Section~\ref{sec:mcmc}. For each specification, the sampler is run for 3{,}000 burn-in iterations followed by 5{,}000 retained draws. Latent expert signals are sampled at each iteration, and the entry and exit operators described in Section~\ref{sec:entry_exit} ensure that posterior inference remains coherent under the time-varying composition of the active set $A_t$.

Prior specifications follow standard choices in the BPS literature \citep[e.g.,][]{mcalinn2019dynamic,mcalinn2021mixed}. Specifically, conditional on the observation variance $v_0$, the initial synthesis coefficients satisfy $\btheta_{0}|v_{0}\sim N(\m_{0},(v_{0}/s_{0})\C_0)$ with $\mathbf{m}_0 = (1/J)\mathbf{1}'$ and $\mathbf{C}_0 = 10^{-4}\,I$. The observation variance prior is $1/v_{0}\sim G(n_{0}/2,n_{0}s_{0}/2)$ with $n_{0}=5,s_{0}=0.01$. Discount factors are set to $(d,\beta)=(0.99,0.9)$, where $d$ discounts state evolution and $\beta$ discounts stochastic volatility. Both prior and discount factor specifications follow \cite{mcalinn2019dynamic}.

{We set the latent correlation parameter in $M_c$ to $M_{\mathrm{corr}}=0.99$ and set the variance for entering/exiting experts to $1$.} For the entry $\theta_{A_{t-1}^c}$, we assume a Normal distribution with mean in $\{0,1/J\}$ (or, in the case of re-entry, the last available parameter) and variance $1$, consistent with the prior on the initial experts. {These choices span pessimistic, neutral, and information-preserving entry scenarios.} Forecasts are computed by drawing from the one-year-ahead predictive distribution and applying the appropriate future entry/exit operators.
Supplementary Material~\ref{app:robustspecification} reports results across the full grid of latent correlation settings $\rho \in \{0, 0.5, 0.9, 0.99\}$; the qualitative conclusions are robust to this choice.

\subsection{Evaluation metrics}
We assess both point and density forecasting performance. Point accuracy is measured using the root mean squared forecast error (RMSE),
\[
\text{RMSE}
= \sqrt{\frac{1}{T}
  \sum_{t=1}^{T}
  (y_{t} - \hat{y}_{t})^2}.
\]
{For ease of interpretation, RMSE values are reported relative to the equal-weight (EW) benchmark, so that values below one indicate an improvement over equal weighting. }

Density forecast accuracy is evaluated using the log predictive density ratio (LPDR) relative to the equal-weight (EW) benchmark,
\[
\text{LPDR}
= 
  \sum_{t=1}^{T}
  \Big[
    \log p(y_{t}| D_t) -
    \log p_{\text{EW}}(y_{t}| D_t)
  \Big],
\]
where $D_t$ denotes the information set available at time $t$. Positive LPDR values favor the method under consideration relative to equal weighting. All results are based on next-quarter year-ahead inflation forecast, where forecasts are constructed recursively to mimic real-time implementation.

\subsection{Overall forecasting performance} \label{subsec:fullsample}

\begin{figure}[!t]  
    \centering

\setlength{\tabcolsep}{0\textwidth}
\begin{tabular}{cc}
\small{Cumulative root mean squared forecast error (RMSE).} & \small{Cumulative log predictive density ratio (LPDR).}\\
    \includegraphics[width=.49\linewidth]{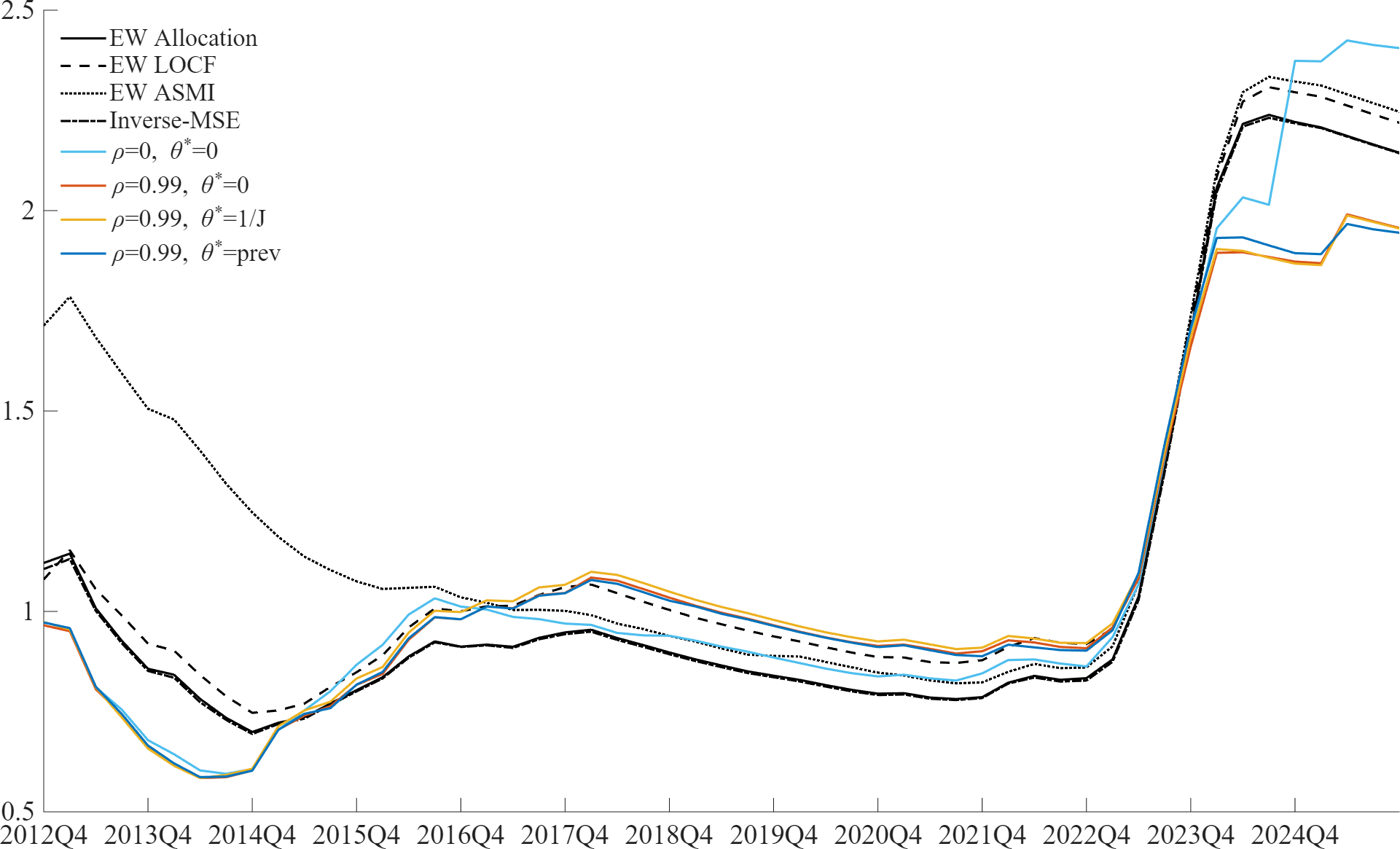} &
    \includegraphics[width=.49\linewidth]{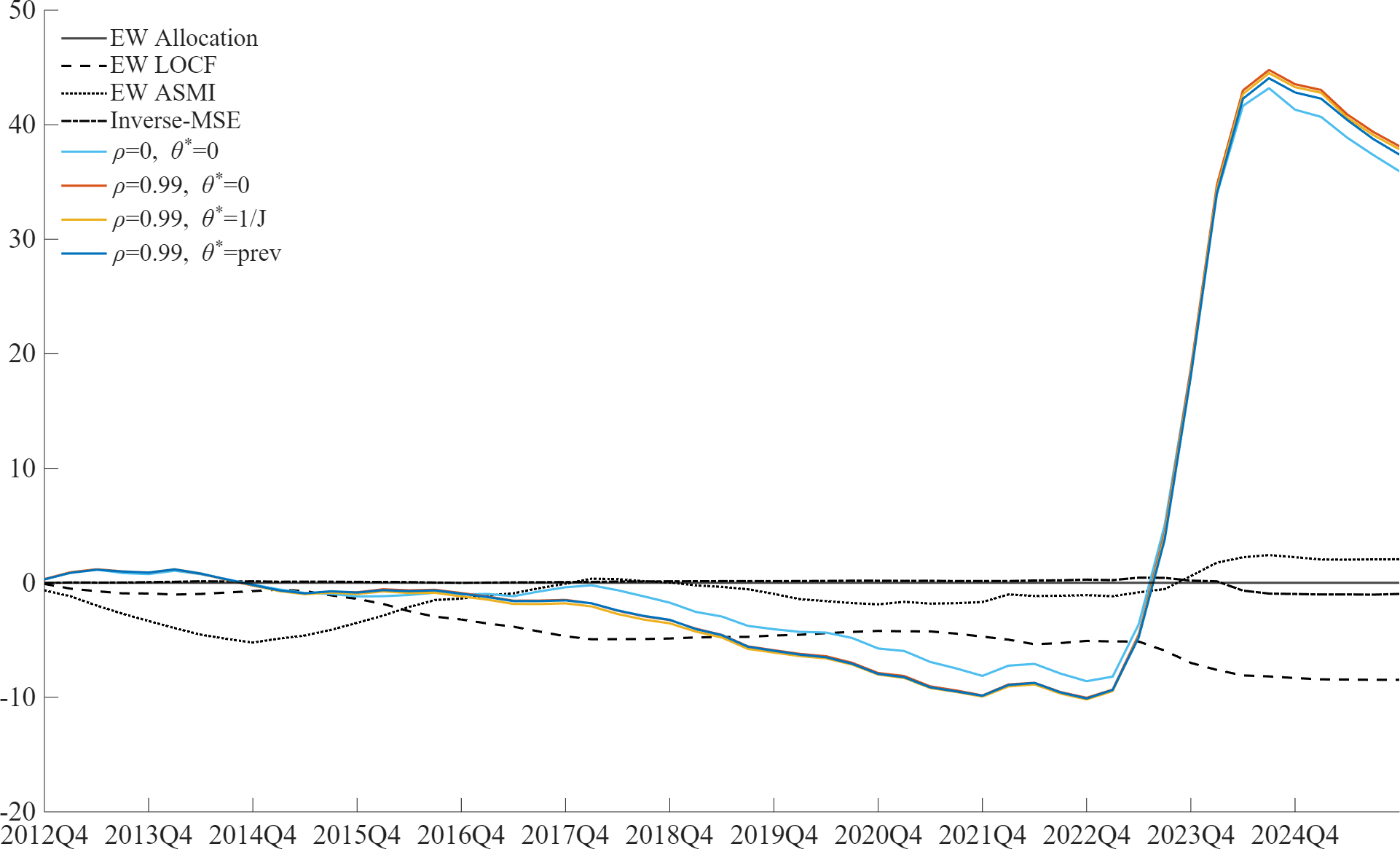} \\   
\end{tabular}
    



    \caption{One-year-ahead predictive accuracy of the proposed coherent Bayesian method 
    compared with equal weights. Panel (a) tracks the evolution of RMSE in levels, and panel (b) 
    reports the cumulative LPDR relative to the equal-weight benchmark.}
    \label{fig:mse_lpdr}
\end{figure}

Table~\ref{tab:results} and Figure~\ref{fig:mse_lpdr} summarize one-year-ahead (four-quarter-ahead) forecasting performance over the full evaluation period. Two messages emerge. First, coherent BPS delivers the strongest {density} forecasts by a wide margin, with large cumulative LPDR gains relative to equal weights. Second, {point} accuracy differences are comparatively modest for most of the sample, but become economically meaningful in high-volatility episodes-- precisely when the panel is most irregular and forecaster turnover is most acute.
The results under alternative prior specifications are given in Supplementary Material~\ref{sec:add}.

\begin{table}[t]
\centering
\small
\caption{Full-sample forecasting performance over the evaluation period.
RMSE measures one-year-ahead point accuracy, relative to EW; LPDR is the log predictive
density ratio against the EW benchmark (positive is better). The latent correlation is set to $0.99$}
\begin{tabular}{C{.1\textwidth}C{.1\textwidth}C{.1\textwidth} C{.1\textwidth} C{.1\textwidth} C{.1\textwidth} C{.1\textwidth} C{.1\textwidth} } \hline \hline
	&	EW &EW LOCF 	&	 EW ASMI 	&	 Inverse-MSE	&	$\theta^*=0$ 	&	$\theta^*=1/J$	&	$\theta^*=prev$	\\ \hline
RMSE 	&1.00	&1.03	&	1.05	&	1.13	&	0.92	&	0.92	&	0.93	\\
LPDR 	&0.00	&-8.45	&	2.18	&	-0.86	&	33.07	&	32.81	&	31.94	\\\hline
\end{tabular} 
\label{tab:results}
\end{table}

\paragraph{Point accuracy (RMSE).}
Panel~(a) of Figure~\ref{fig:mse_lpdr} shows that all approaches track each other closely during long tranquil stretches, but separate in periods when the information set becomes sparse or shifts abruptly. The coherent BPS specifications with high latent correlation (e.g., $\rho=0.99$) achieve the lowest or near-lowest cumulative RMSE over the full sample, and the advantage is driven primarily by the most volatile subperiods. Among the coherent variants, $\theta^*=0$ and $\theta^*=1/J$ are essentially indistinguishable in RMSE, while $\theta^*=\text{prev}$ is modestly worse, consistent with propagating stale coefficient information across long gaps between participation.

Imputation-based competitors (LOCF and ASMI) do not improve point accuracy and can be worse than simple equal weighting, reflecting a basic mismatch: missingness here is largely \emph{structural} (entry/exit and sporadic reporting), not short-lived measurement loss. Inverse-MSE weighting can match equal weights in calmer periods but remains sensitive to abrupt changes in the active set, producing less stable performance when turnover is high.

\paragraph{Density accuracy (LPDR).}
Panel~(b) of Figure~\ref{fig:mse_lpdr} is the clearest discriminator. Coherent BPS produces very large positive cumulative LPDR relative to equal weights, indicating substantially better calibrated predictive distributions and systematically higher probability mass assigned to realized outcomes. By contrast, LOCF is strongly negative, and inverse-MSE is near zero or negative, reflecting overconfidence and poor uncertainty propagation when the forecaster set changes. ASMI can be mildly positive, but its gains are small relative to coherent BPS.

A notable feature is that density gains are {not uniform over time}: coherent BPS can lag in the middle of the sample (cumulative LPDR drawdowns), but it dominates sharply during the inflation surge, when the realized outcomes move into regions that linear pools and shrinkage heuristics systematically underweight. This is exactly the setting where coherence-based adjustment matters most: the model expands and reallocates uncertainty in response to changing participation and elevated dispersion, rather than implicitly treating the panel as stable.

\paragraph{Temporal profile and regime dependence.}
Performance differences are time-localized. Early in the evaluation window, methods are broadly similar. In the middle of the sample, when participation is especially uneven, coherent BPS is more conservative and can temporarily lose ground in cumulative metrics. The ranking reverses decisively during the COVID-19 period and inflation surge: coherent BPS generates the largest gains in both RMSE and LPDR, and these gains dominate the full-sample totals.

\paragraph{Interpretation.}
In a sporadic panel, the central difficulty is not merely choosing weights, but maintaining a coherent mapping from an evolving, partially observed set of agent forecasts to a predictive distribution for the target. Methods that impute missing forecasts or reweight mechanically can look competitive in benign periods, but they do not propagate uncertainty correctly when the panel composition changes. Coherent BPS delivers its largest improvements during high-volatility regimes with irregular participation, where coherent adjustment prevents spurious jumps in the latent state while still allowing the model to adapt quickly when the signal content of the panel changes.

\paragraph{Why density gains exceed point gains.}
The large gap between density improvement (LPDR) and point improvement (RMSE) reflects a fundamental asymmetry in what coherent updating corrects. Point forecasts depend primarily on the synthesis coefficients $\theta_t$, which all methods can estimate reasonably well when participation is only mildly irregular. Density forecasts, however, depend critically on the predictive variance, which in turn depends on how uncertainty is propagated when the forecaster set changes. When a forecaster exits, naive methods either drop that forecaster's contribution entirely-- loss of variance-- or impute a stale density-- artificial variance from outdated information. Neither approach correctly quantifies how much additional uncertainty arises from losing an information source. The coherent exit operator, by contrast, explicitly inflates the predictive variance through the marginalization in equation \eqref{eq:exit_var}: the additional terms add variance precisely because the exiting forecaster's latent state is now unobserved. Symmetrically, coherent entry adjusts variance downward as new information arrives. This variance accounting is invisible in point metrics but decisive for density calibration, explaining why the LPDR gains are an order of magnitude larger than the RMSE gains and why these gains concentrate in high-turnover periods when the variance adjustments are most active.

\subsection{Dynamics of synthesis weights}

\begin{figure}[!t]
    \centering

    \begin{subfigure}{0.48\linewidth}
        \centering
        \includegraphics[width=\linewidth]{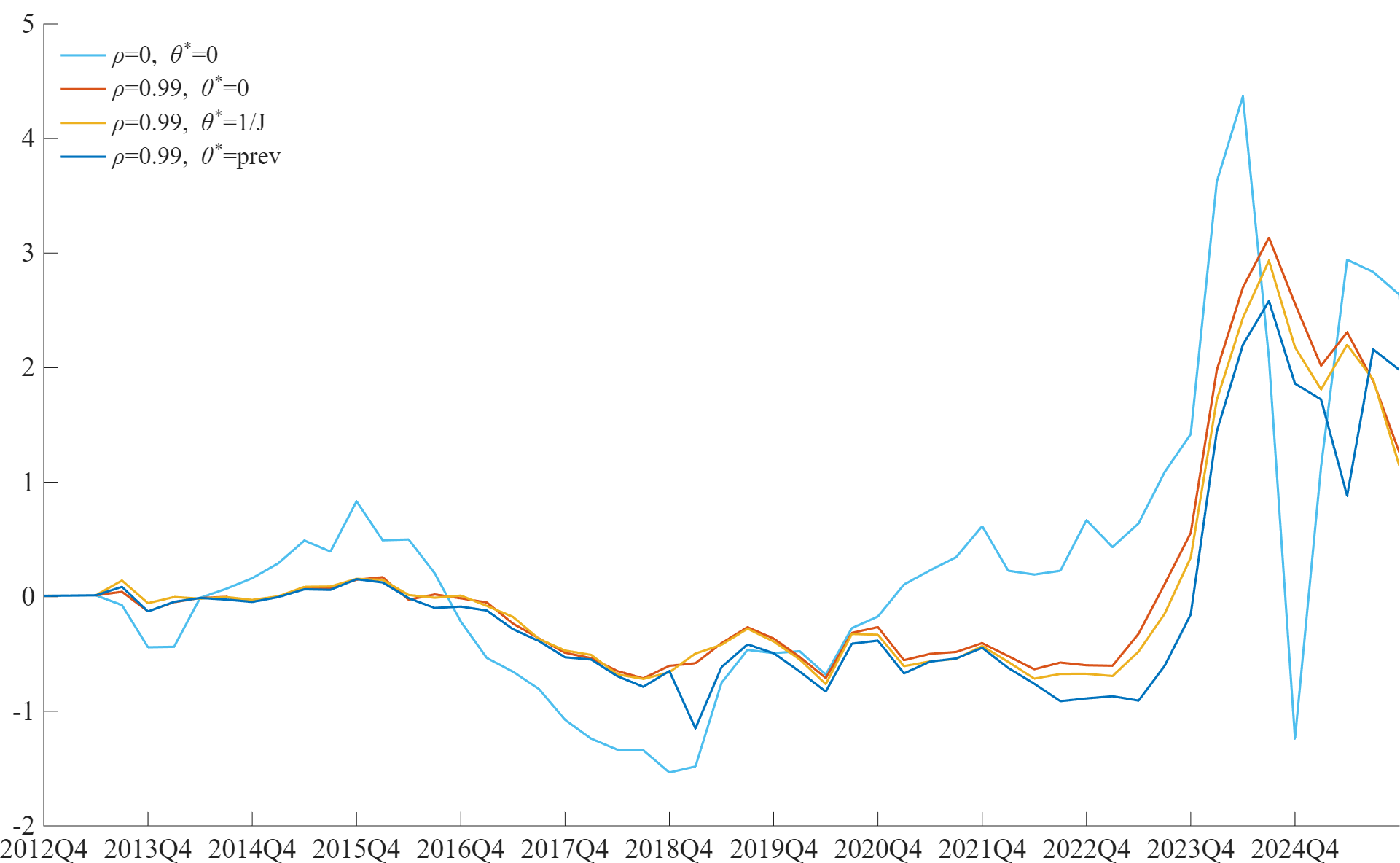}
        \caption{Intercept paths of BPS.}
    \end{subfigure}
    \hfill
    
    \vspace{0.8em}

    \begin{subfigure}{0.48\linewidth}
        \centering
        \includegraphics[width=\linewidth]{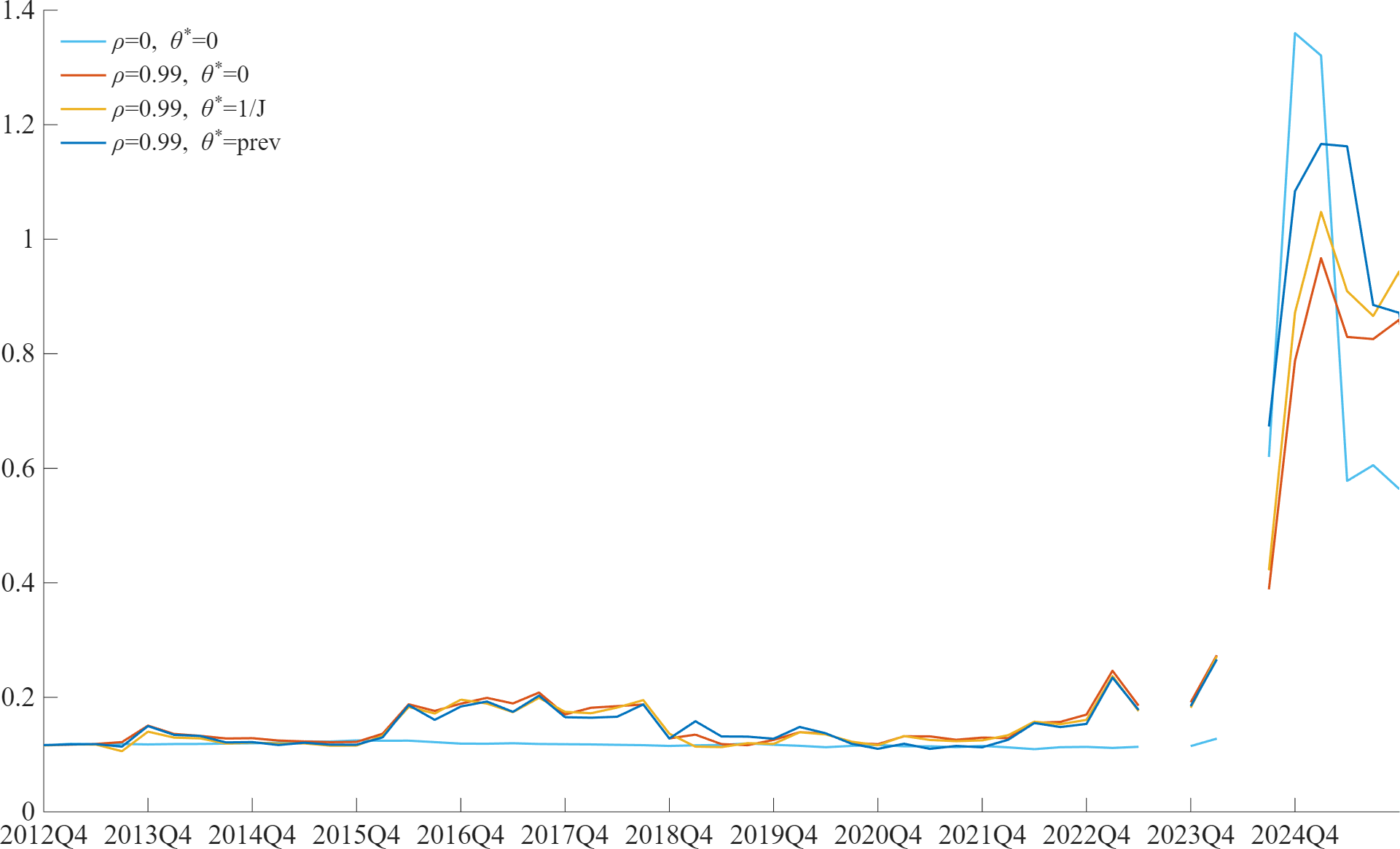}
        \caption{Selected coefficient paths, $\rho=0.99$.}
    \end{subfigure}
    \hfill
    \begin{subfigure}{0.48\linewidth}
        \centering
        \includegraphics[width=\linewidth]{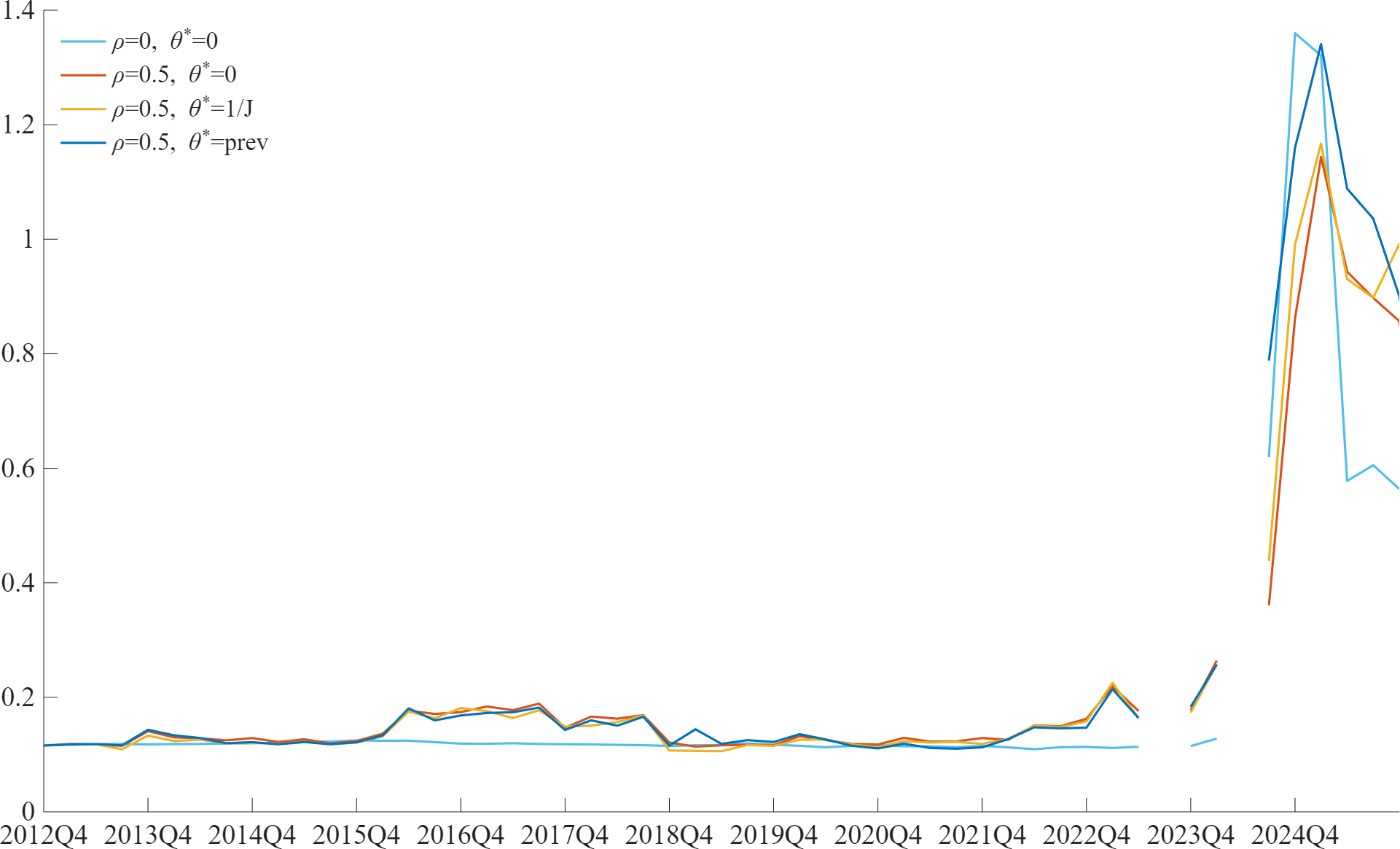}
        \caption{Selected coefficient paths, $\rho=0.5$.}
    \end{subfigure}

    \caption{Evolution of posterior coefficients under different correlation and prior 
    specifications. The coherent BPS model yields smooth, interpretable paths even in the 
    presence of frequent entry and exit.} 
    \label{fig:coefpaths}
\end{figure}

Figure~\ref{fig:coefpaths} illustrates the evolution of posterior synthesis coefficients under representative specifications. Panel~(a) shows the intercept path, which captures low-frequency movements in the latent inflation component, while panels~(b)--(c) show coefficient paths for selected forecasters under different latent-correlation settings.

Two features stand out. First, different latent dependence settings (e.g., $\rho=0.99$ vs.\ $\rho=0.5$) do not drastically alter coefficient dynamics; smoothness and interpretability remain stable. The intercept and persistent-forecaster coefficients evolve smoothly through routine entry/exit events, with no spurious breaks for continuing experts. Second, the intercept responds most strongly during macro stress periods. During the inflation surge, the model reallocates weight sharply toward the intercept, consistent with the large LPDR gains in Figure~\ref{fig:mse_lpdr}. More informative entry priors lead to less allocation to the intercept, as reflected in panels~(b)--(c).

Weights for persistent experts evolve smoothly, reflecting gradual learning about relative forecasting performance. Importantly, entry and exit events do not induce artificial jumps in the intercept or in the coefficients of continuing forecasters because the coherence-based adjustment operators ensure that the global state remains stable even as the composition of $A_t$ changes.

\subsection{Isolated analysis: impact of entry and exit}
\begin{figure}[!t]
    \centering

    \begin{subfigure}[t]{0.49\linewidth}
        \centering
        \includegraphics[width=\linewidth]{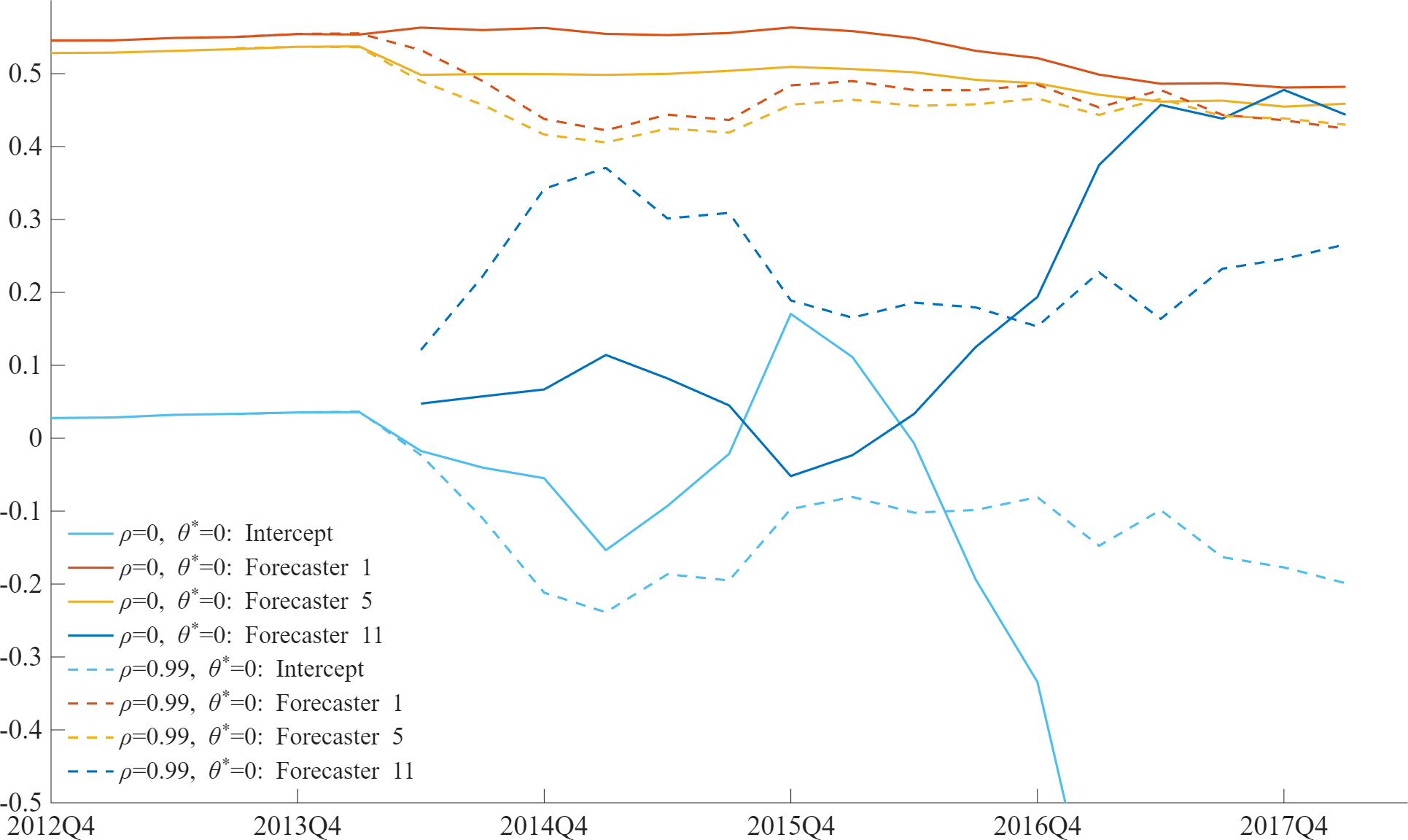}
        \caption{Entry adjustment, prior A.}
    \end{subfigure}\hfill
    \begin{subfigure}[t]{0.49\linewidth}
        \centering
        \includegraphics[width=\linewidth]{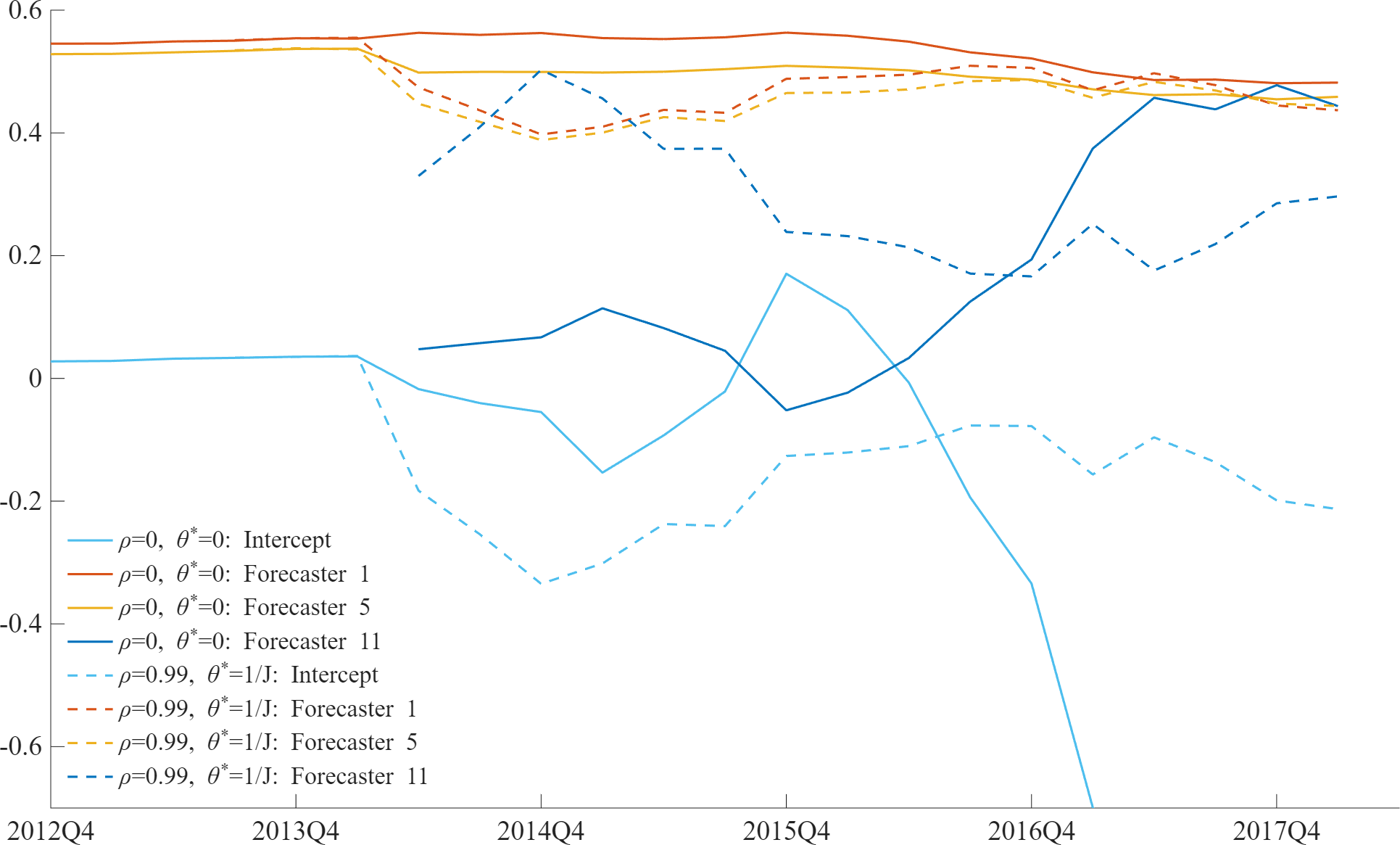}
        \caption{Entry adjustment, prior B.}
    \end{subfigure}

    \vspace{0.8em}

    \begin{subfigure}[t]{0.49\linewidth}
        \centering
        \includegraphics[width=\linewidth]{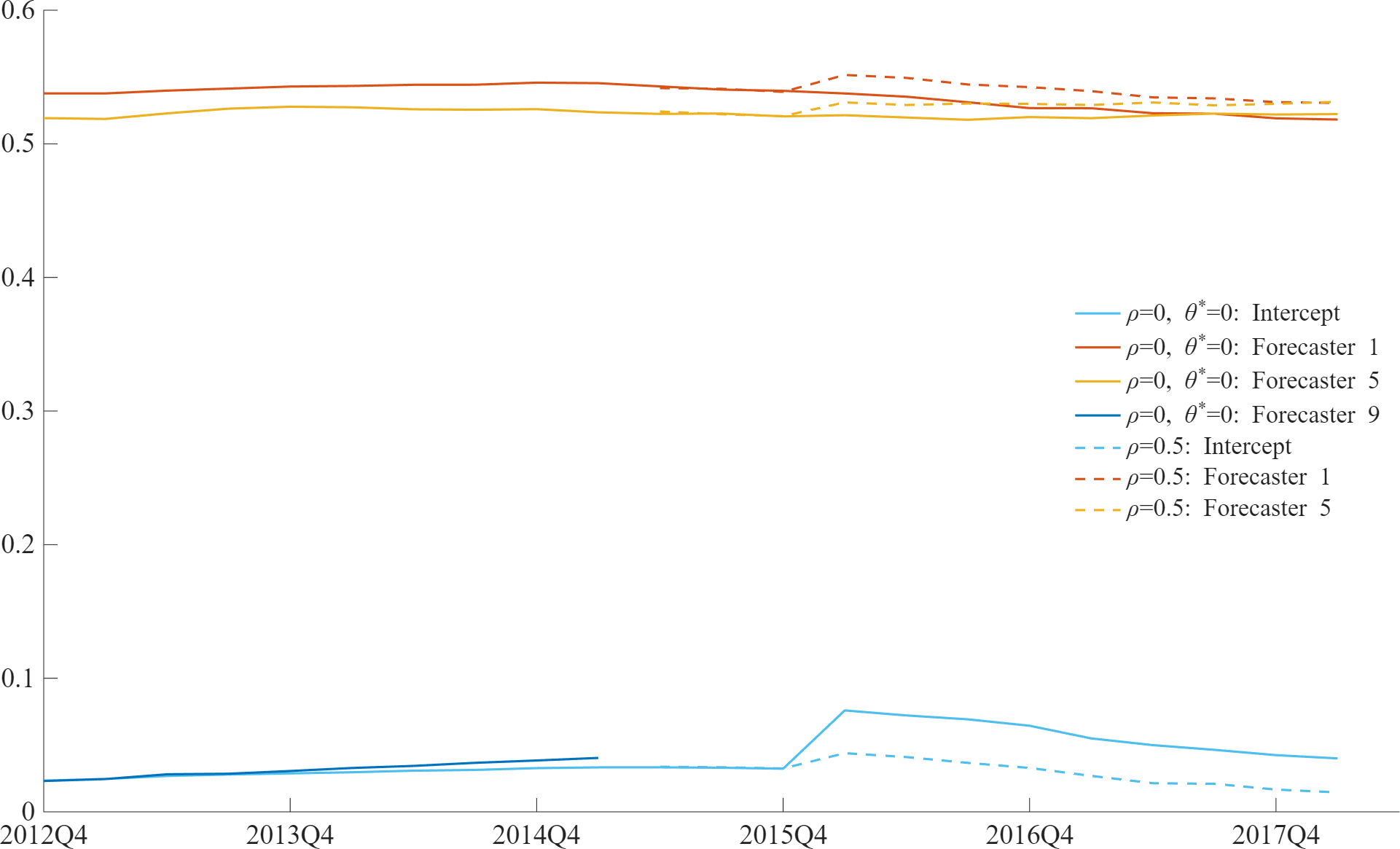}
        \caption{Exit adjustment, prior A.}
    \end{subfigure}\hfill
    \begin{subfigure}[t]{0.49\linewidth}
        \centering
        \includegraphics[width=\linewidth]{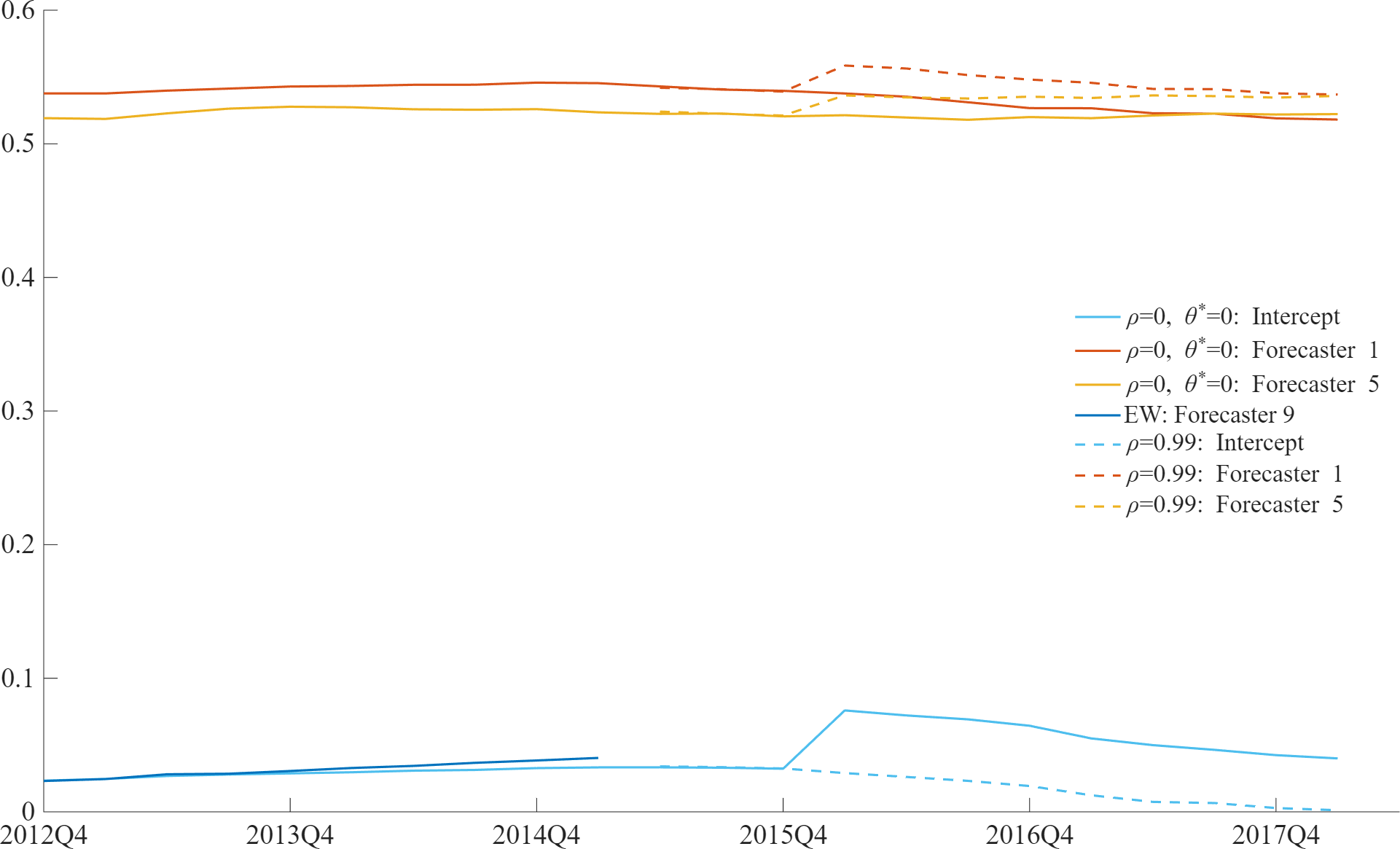}
        \caption{Exit adjustment, prior B.}
    \end{subfigure}

    \caption{Illustration of the coherent adjustment to posterior coefficients during entry and exit events.}
    \label{fig:entry-exit}
\end{figure}

Figure~\ref{fig:entry-exit} provides a focused illustration of the entry and exit mechanisms, zooming in on short windows around single changes in the expert set. We consider a reduced panel consisting of two persistently active experts (red and yellow lines) and one expert that either enters or exits (dark blue line), allowing the adjustment mechanism to be examined in isolation. 

During entry events, the coherent BPS update moves the intercept and associated coefficients smoothly toward values consistent with the expanded panel, without generating spurious spikes. Similarly, during exit events, the contributions of departing experts phase out without inducing discontinuities in the remaining coefficients. Naive alternatives that simply drop missing forecasters or reset their coefficients to zero would produce visible breaks in these paths.

%
%
\section{Discussion} \label{sec:discussion}
The empirical results underscore that when a forecast survey panel is unstable, the mechanics of how entry, exit, and missing submissions are handled can materially affect both real-time predictive uncertainty and retrospective evaluation. Procedures that drop and renormalize, or that impute missing submissions, implicitly alter the conditioning information set in ways that can induce artificial movements in combined predictive moments. The coherence-based updating developed here addresses this by treating missing forecasts as unobserved latent predictive states rather than deleted contributors, yielding smooth predictive evolution as the panel composition changes.

The main practical benefits are threefold. First, the resulting predictive distribution is less sensitive to administrative variation in who happens to report; changes are driven primarily by reported information and posterior learning. Second, because entry and exit are absorbed through marginalization and conditional projection rather than ad hoc weight reallocation, changes in synthesis coefficients can be interpreted as genuine shifts in relative predictive value. Third, the method can be implemented within standard filtering pipelines and does not require specifying a structural participation model. When panel instability is mild, simple pooling may be adequate; when entry, exit, and long participation gaps are common, a coherent updating rule prevents the combined predictive distribution from inheriting mechanical discontinuities at precisely the times when uncertainty quantification is most salient.

The approach does not attempt to model why forecasters miss rounds. Instead, it targets a narrower but operationally central objective: given the subset of densities observed at time $t$, update the combined predictive distribution consistently with a single Bayesian joint model. This coherence-first stance is appropriate when the goal is a deployable aggregation rule that avoids artifacts rather than causal interpretation of participation itself. The approach inherits limitations from the underlying dynamic synthesis framework: the Gaussian specification may be restrictive for strongly non-Gaussian predictive states, and entry events require prior specification for entering coefficients. In practice, sensitivity to these priors is typically localized to short windows around entry events.

Several extensions are immediate: multivariate and multi-horizon forecasting, hierarchical priors for grouped forecasters, alternative (non-Gaussian) synthesis families, and application to other sporadic panels in corporate planning, energy-load forecasting, and epidemiological forecast hubs. The empirical lesson is that irregular participation is a first-order feature, and combination rules should be designed with it explicitly in mind.

%
%
\section{Conclusion}\label{sec:conclusion}

Professional forecaster surveys are a central input for inflation monitoring and policy communication, yet their panels are inherently sporadic. Entry, exit, and intermittent participation imply that the set of available density forecasts changes over time, so standard pooling rules can conflate genuine shifts in beliefs about the economy with mechanical changes in panel composition. This paper develops a coherent Bayesian predictive synthesis approach for such environments by associating each forecaster with a latent predictive state that exists regardless of whether a forecast is observed, and by updating the combined predictive distribution through Bayesian conditioning under partial observation.

Applied to inflation density forecasts from a professional forecaster survey, the proposed approach improves both predictive performance and interpretability relative to commonly used operational baselines. The gains are most pronounced during periods of elevated panel instability-- precisely when reliable and stable measures of uncertainty are most valuable for policy analysis. More broadly, the results underscore a general point: when the objective is a deployable combined predictive distribution, handling forecaster turnover and missingness is not a secondary data issue but an integral part of the statistical problem. Enforcing coherence provides a principled and practical alternative to ad hoc renormalization or imputation, yielding forecast combination systems that remain stable, interpretable, and informative under the sporadic participation patterns that characterize real-world forecasting panels.

\singlespacing
\bibliographystyle{chicago}
\bibliography{reference}

\newpage
\doublespacing

%
\setcounter{equation}{0}
\setcounter{section}{0}
\setcounter{table}{0}
\setcounter{page}{1}
\renewcommand{\thesection}{S\arabic{section}}
\renewcommand{\theequation}{S\arabic{equation}}
\renewcommand{\thetable}{S\arabic{table}}

\vspace{1cm}
\begin{center}
{\LARGE
{\bf Supplementary Material for ``Predictive Synthesis under Sporadic Participation: Evidence from Inflation Density Surveys"}
}
\end{center}

%
%
\section{Posterior Computation and Filtering Algorithm}\label{sec:mcmc}

This supplement describes the exact posterior simulation scheme used in our empirical analysis.

\subsection{Model components used in computation}

At each quarter $t=1{:}T$, conditional on latent expert states $\x_t=(x_{1,t},\ldots,x_{J,t})'$ and synthesis
coefficients $\btheta_t=(\theta_{0,t},\theta_{1,t},\ldots,\theta_{J,t})'$, the synthesis likelihood is the
Gaussian DLM
\[
y_t = F_t'\btheta_t + \nu_t,
\qquad
F_t=(1,x_{1,t},\ldots,x_{J,t})',
\qquad
\nu_t\sim N(0,v_t),
\]
with coefficients corresponding to inactive experts deterministically set to zero.

We represent each reported expert density $h_{j,t}$ by a Student-$t$ form with parameters
(mean $a_{j,t}$, variance $A_{j,t}$, df $n_{j,t}$) and use the standard normal--gamma mixture:
\[
x_{j,t}| \phi_{j,t} \sim N\!\bigl(a_{j,t},\, \phi_{j,t}A_{j,t}\bigr),
\qquad
\phi_{j,t}\sim \text{Gamma-mix}(n_{j,t}),
\]
where $\phi_{j,t}$ is sampled via the conjugate update in Step~\ref{step:phi} below.

Let $A_t\subseteq\{1,\ldots,J\}$ be the active set at time $t$ (indicator vector $\mathrm{idx}(t,\cdot)$).
Define entry/exit sets between $t-1$ and $t$:
\[
\mathcal{X}_t=\{j:\mathrm{idx}_{t-1,j}=1,\ \mathrm{idx}_{t,j}=0\},\qquad
\mathcal{E}_t=\{j:\mathrm{idx}_{t-1,j}=0,\ \mathrm{idx}_{t,j}=1\}.
\]

\subsection{Discount-state forward filter}

Conditional on a fixed draw of $\x_t$ (within an MCMC iteration), the state evolution for $\btheta_t$ is handled
via discounting. If $(m_t,C_t)$ denotes the filtered mean/covariance \emph{after observing $y_{t-1}$} and
before processing time $t$, the time-$t$ prior is
\[
a_t=m_t,\qquad R_t=C_t/d,
\]
where $d\in(0,1]$ is the state discount factor. Volatility is updated using the standard discount stochastic
volatility recursion (implemented through $(n_t,s_t)$):
\[
n_{t+1}=\beta n_t+1,\qquad
s_{t+1}=s_t\cdot \frac{\beta n_t + e_t^2/q_t}{n_{t+1}},
\]
where $\beta\in(0,1]$ is the volatility discount factor, $e_t=y_t-f_t$, and $q_t$ is the one-step forecast
variance defined below.

Given $(a_t,R_t,s_t,n_t)$, the usual one-step forecast and update are
\[
f_t=F_t'a_t,\qquad q_t=F_t'R_tF_t+s_t,\qquad A_t=R_tF_t/q_t,
\]
\[
m_{t+1}=a_t + A_t(y_t-f_t),\qquad
C_{t+1}= r_t\bigl(R_t-q_tA_tA_t'\bigr),
\qquad
r_t=\frac{\beta n_t + e_t^2/q_t}{n_{t+1}}.
\]

\subsection{Coherent prior adjustment for exit and entry}

When the active set changes between $t-1$ and $t$, coherence requires conditioning on the observed subset
without renormalization or deletion. In computation, we implement this by {transporting the Gaussian prior}
$N(a_t,R_t)$ to an adjusted prior $N(\tilde a_t,\tilde R_t)$ that corresponds to the coherence-implied
marginal/extension, and then using $(\tilde a_t,\tilde R_t)$ to compute $f_t,q_t$ and update with $y_t$.

\paragraph{Latent-state covariance used by the operators.}
To construct the regression operators linking entering/exiting experts to continuing experts, we use a
working joint covariance for $\x_t$:
\[
\Sigma_t = D_t M_c D_t,
\qquad
D_t=\mathrm{diag}\!\bigl(\sqrt{\phi_{1,t}A_{1,t}},\ldots,\sqrt{\phi_{J,t}A_{J,t}}\bigr),
\]
where $M_c$ is a correlation matrix with off-diagonals set to the scalar $M_{\mathrm{corr}}$ (and ones on the
diagonal). When some $A_{j,t}$ must be assigned for experts not used in a given operator call (e.g., to define
$\Sigma_t$ on an exit/entry set), we follow the empirical default in the code: replace missing/zero entries by
the mean of the nonzero variances at time $t$.

\paragraph{Monte Carlo transport.}
Both exit and entry operators are implemented by:
(i) drawing $S$ samples $\theta^{(s)}\sim N(a_t,R_t)$ restricted to the relevant active set,
(ii) applying a linear map $\theta^{(s)}\mapsto \theta^{(s)\,*}$,
(iii) setting $(\tilde a_t,\tilde R_t)$ to the sample mean/covariance of $\{\theta^{(s)\,*}\}$ and enforcing
the current activity mask (inactive coefficients set to zero).
In our implementation $S=1000$.

\subsubsection{Exit operator}

If $\mathcal{X}_t\neq\emptyset$, let $\mathcal{C}_t=\{1,\ldots,J\}\setminus \mathcal{X}_t$ denote
non-exiting indices. Define the regression matrix
\[
B^{\text{exit}}_t = \Sigma_{\mathcal{X}_t,\mathcal{C}_t,t}\,
\Sigma_{\mathcal{C}_t,\mathcal{C}_t,t}^{-1},
\]
and let $\mu_t=(a_{1,t},\ldots,a_{J,t})'$ be the vector of expert forecast means at time $t$.
For each prior draw $\btheta^{(s)}=(\theta_0^{(s)},\theta_1^{(s)},\ldots,\theta_J^{(s)})'$, define the exit
transport (written in block form, matching the code):
\[
\theta_0^{(s)\,*}
=
\theta_0^{(s)}
+
\theta_{\mathcal{X}_t}^{(s)\,'}
\Bigl(\mu_{\mathcal{X}_t,t}-B^{\text{exit}}_t\,\mu_{\mathcal{C}_t,t}\Bigr),
\]
\[
\theta_{1:J}^{(s)\,*}
=
\theta_{1:J}^{(s)}
+
\theta_{\mathcal{X}_t}^{(s)\,'}\,\bar B^{\text{exit}}_t,
\]
where $\bar B^{\text{exit}}_t$ is the padded version of $B^{\text{exit}}_t$ with zeros in the columns
corresponding to $\mathcal{X}_t$ and $B^{\text{exit}}_t$ placed in the columns for $\mathcal{C}_t$.
Finally, enforce the current activity mask at time $t$ by setting $\btheta_{j}^{(s)\,*}=0$ for all $j\notin A_t$.
The adjusted prior moments $(\tilde a_t,\tilde R_t)$ are the sample mean/covariance of $\btheta^{(s)\,*}$.

\subsubsection{Entry operator}

If $\mathcal{E}_t\neq\emptyset$, let $\mathcal{C}_t=\{1,\ldots,J\}\setminus \mathcal{E}_t$ denote
non-entering indices. Define
\[
B^{\text{ent}}_t = \Sigma_{\mathcal{E}_t,\mathcal{C}_t,t}\,
\Sigma_{\mathcal{C}_t,\mathcal{C}_t,t}^{-1}.
\]
Entry requires specifying a prior for the new coefficients. In the implementation, this is done by:
(i) resetting all covariances involving entering coefficients to zero,
(ii) assigning their marginal variances to a base value, and
(iii) assigning entering prior means.
Denote the resulting ``reset'' prior by $N(a_t^{(0)},R_t^{(0)})$.

Then draw $\btheta^{(s)}\sim N(a_t^{(0)},R_t^{(0)})$ (restricted to current active set), and apply the entry
transport
\[
\theta_0^{(s)\,*}
=
\theta_0^{(s)}
-
\theta_{\mathcal{E}_t}^{(s)\,'}
\Bigl(\mu_{\mathcal{E}_t,t}-B^{\text{ent}}_t\,\mu_{\mathcal{C}_t,t}\Bigr),
\]
\[
\theta_{\mathcal{C}_t}^{(s)\,*}
=
\theta_{\mathcal{C}_t}^{(s)}
-
\theta_{\mathcal{E}_t}^{(s)\,'}\,B^{\text{ent}}_t,
\qquad
\theta_{\mathcal{E}_t}^{(s)\,*}=\theta_{\mathcal{E}_t}^{(s)}.
\]
As in exit, we enforce the activity mask at time $t$ and set $(\tilde a_t,\tilde R_t)$ to the sample
mean/covariance of $\btheta^{(s)\,*}$.

\paragraph{Ordering.}
When both exit and entry occur at the same $t$, we apply \emph{exit first, then entry}, both on the
time-$t$ prior, and only then compute $f_t,q_t$ and update with $y_t$. 

\subsection{Full MCMC algorithm}\label{subsec:full_mcmc}

Algorithm~\ref{alg:sporadic_bps_mcmc} summarizes the complete Gibbs/FFBS scheme.
It conditions on the previous latent-state draw $\x^{(i-1)}$ when filtering $\btheta$, then conditions on the
new $\btheta^{(i)}$ when drawing $\x^{(i)}$.

\begin{algorithm}[t]
\caption{Sporadic BPS MCMC with coherent prior adjustment}\label{alg:sporadic_bps_mcmc}
\begin{algorithmic}[1]
\State \textbf{Inputs:} $\{y_t\}_{t=1}^T$; expert parameters $\{a_{j,t},A_{j,t},n_{j,t}\}$; activity indicators
$\{\mathrm{idx}_{t,j}\}$; discounts $(d,\beta)$; priors $(m_0,C_0,n_0,s_0)$; hyperparameters $(M_{\mathrm{corr}},\theta_m,\theta_C)$.
\State Precompute entry/exit indicators $\mathcal{E}_t,\mathcal{X}_t$ from $\mathrm{idx}$.
\State Initialize $\{\x_t^{(0)},\phi_t^{(0)}\}$ by sampling from the expert mixture representation.
\For{$i=1,\ldots,I$}
    \State \textbf{(Optional)} Select $(d,\beta)$ by maximizing the one-step-ahead log predictive likelihood
    over a candidate grid (singleton in our empirical runs).
    \State Initialize $(m_1,C_1,n_1,s_1)=(m_0,C_0,n_0,s_0)$.
    \For{$t=1,\ldots,T$}
        \State Form $F_t=(1,x_{1,t}^{(i-1)},\ldots,x_{J,t}^{(i-1)})'$ and apply the activity mask (inactive coefficients fixed at $0$).
        \State Compute the discounted prior $(a_t,R_t)$ from $(m_t,C_t)$.
        \State \textbf{Coherent prior adjustment:}
        \If{$\mathcal{X}_t\neq\emptyset$} apply the \emph{exit} operator to $(a_t,R_t)$ via Monte Carlo transport to obtain $(\tilde a_t,\tilde R_t)$. \Else set $(\tilde a_t,\tilde R_t)=(a_t,R_t)$. \EndIf
        \If{$\mathcal{E}_t\neq\emptyset$} apply the \emph{entry} operator to $(\tilde a_t,\tilde R_t)$ via the reset+transport step to obtain final prior moments $(\tilde a_t,\tilde R_t)$. \EndIf
        \State Using $(\tilde a_t,\tilde R_t)$, compute $f_t,q_t$ and update $(m_{t+1},C_{t+1},n_{t+1},s_{t+1})$ with $y_t$.
    \EndFor
    \State \textbf{Backward sampling (FFBS):} draw $v_{1:T}^{(i)}$ and $\theta_{1:T}^{(i)}$ from the smoothing distribution implied by the discount DLM, enforcing the activity mask at each $t$.
    \For{$t=1,\ldots,T$}
        \State Draw $\x_t^{(i)}$ from its conditional Gaussian posterior given $(\btheta_t^{(i)},v_t^{(i)},y_t)$ (conjugate update).
        \State \label{step:phi} Draw $\phi_{t}^{(i)}$ from its conditional gamma-mixture update given $\x_t^{(i)}$ (Student-$t$ mixture representation).
    \EndFor
\EndFor
\State \textbf{Output:} posterior draws after burn-in; one-step-ahead predictive parameters $(a_{T+1},R_{T+1},v_{T+1},n_{T+1})$ computed from the final filtered state, with the same exit/entry logic applied to the $T{+}1$ active set.
\end{algorithmic}
\end{algorithm}

%
%
\section{Robustness analysis}\label{sec:add}

This appendix reports robustness checks and extensions of the main empirical results. We document forecasting performance across the full grid of coherence and entry-prior specifications considered in the sensitivity analysis, as well as under an expanded forecaster set. These additional analyses confirm that the paper's qualitative conclusions are robust to alternative specifications and panel definitions.

\subsection{Different model specifications}\label{app:robustspecification}

Table~\ref{tab:results_full} reports the complete set of specifications considered in the robustness grid. Each row corresponds to one combination of (i) the correlation setting governing cross-forecaster dependence in the latent forecast-state prior (parametrized by $\rho$), and (ii) the entry prior option for newly active forecasters (e.g., mean initialization choice). For each specification, we re-run the full real-time forecasting evaluation and recompute all benchmarks on the same information set, so that any differences reflect only the modeling choice being varied rather than changes in data availability or evaluation design.

Overall, the results are consistent with those in Section \ref{subsec:fullsample}: for point forecasting, the RMSEs are on par or better than the EW benchmark; for density forecasting, LPDRs are uniformly and substantially higher than EW. Moreover, a systematic pattern emerges in the robustness grid: specifications with less informative entrant prior means tend to perform better. This suggests that aggressively incorporating information from newly entering forecasters can be detrimental on average, and that agnostic priors provide a useful form of regularization at entry

\begin{table}[t]
\centering
\caption{Full-sample forecasting performance over the evaluation period.
RMSE measures one-year-ahead point accuracy; LPDR is the log predictive
density ratio against the equal-weight benchmark (positive is better). Target variable: inflation rate. $J=16$}.
\small
\begin{tabular}{lllll}
 & EW & EW LOCF & EW ASMI & Inverse-MSE \\ \hline
RMSE & 2.0690 & 2.1410 & 2.1672 & 2.0676 \\
LPDR & - & -8.44 & 2.18 & -0.85 \\ \hline
 &  &  &  &  \\
 & \multicolumn{4}{c}{$\theta^*=0$} \\ \cline{2-5} 
 & $\rho=0$ & $\rho=0.5$ & $\rho=0.9$ & $\rho=0.99$ \\ \hline
RMSE & 2.3901 & 1.9304 & 1.9171 & 1.9276 \\
LPDR & 30.09 & 32.04 & 32.23 & 31.94 \\ \hline
 &  &  &  &  \\
 & \multicolumn{4}{c}{$\theta^*=1/J$} \\ \cline{2-5} 
 & $\rho=0$ & $\rho=0.5$ & $\rho=0.9$ & $\rho=0.99$ \\ \hline
RMSE & 2.4016 & 1.9084 & 1.9094 & 1.9078 \\
LPDR & 30.17 & 32.81 & 32.91 & 32.82 \\ \hline
 &  &  &  &  \\
 & \multicolumn{4}{c}{$\theta^*=prev$} \\ \cline{2-5} 
 & $\rho=0$ & $\rho=0.5$ & $\rho=0.9$ & $\rho=0.99$ \\ \hline
RMSE & 2.3414 & 1.9097 & 1.9047 & 1.9073 \\
LPDR & 30.70 & 32.82 & 33.17 & 33.07 \\ \hline
\end{tabular}

\label{tab:results_full}
\end{table}

\subsection{Expanding the forecaster set}\label{app:robust_20}

This appendix assesses robustness to enlarging the panel of individual density forecasters. The main text focuses on a curated subset designed to exhibit clear entry and exit and intermittent participation while maintaining a stable core for training. Here we increase the number of included forecasters from $J=16$ to $J=20$ and rerun the full forecast combination exercise, keeping fixed the target alignment (one-year-ahead rolling month), density-to-Normal moment matching, training/evaluation split, BPS prior specification, and the hyperparameter grid over $(\rho,\theta^*)$.

Table~\ref{tab:results_full_20Forecasters} shows that expanding the panel leaves the main conclusions unchanged. In terms of point forecasting, the relative RMSE performance of the synthesis approach remains comparable to that in the baseline $J=16$ specification when evaluated against the EW benchmark. In terms of density forecasting, the synthesis method continues to deliver substantially higher LPDRs than EW pooling across specifications, indicating that its advantages in probabilistic calibration persist as the forecaster set is expanded.

\begin{table}[t]
\centering
\caption{Full-sample forecasting performance over the evaluation period.
RMSE measures one-year-ahead point accuracy; LPDR is the log predictive
density ratio against the equal-weight benchmark (positive is better). Target variable: inflation rate. $J=20$}.
\small
\begin{tabular}{lllll}
 & EW & EW LOCF & EW ASMI & Inverse-MSE \\ \hline
RMSE & 2.0777 & 2.1410 & 2.1672 & 2.0676 \\
LPDR & - & -8.44 & 2.18 & -0.85 \\ \hline
 &  &  &  &  \\
 & \multicolumn{4}{c}{$\theta^*=0$} \\ \cline{2-5} 
 & $\rho=0$ & $\rho=0.5$ & $\rho=0.9$ & $\rho=0.99$ \\ \hline
RMSE & 2.2566 & 1.9976 & 1.9171 & 1.9988 \\
LPDR & 27.17 & 31.19 & 31.14 & 31.10 \\ \hline
 &  &  &  &  \\
 & \multicolumn{4}{c}{$\theta^*=1/J$} \\ \cline{2-5} 
 & $\rho=0$ & $\rho=0.5$ & $\rho=0.9$ & $\rho=0.99$ \\ \hline
RMSE & 2.2916  & 2.0070 & 2.0258 & 2.0166 \\
LPDR & 27.10 & 30.99 & 30.82 & 31.03 \\ \hline
 &  &  &  &  \\
 & \multicolumn{4}{c}{$\theta^*=prev$} \\ \cline{2-5} 
 & $\rho=0$ & $\rho=0.5$ & $\rho=0.9$ & $\rho=0.99$ \\ \hline
RMSE & 2.3076 &  2.0401 &  2.0496 &  2.0451 \\
LPDR & 26.41 & 30.23 & 30.31 & 30.36 \\ \hline
\end{tabular}

\label{tab:results_full_20Forecasters}
\end{table}

%
%

\section{Additional application: one-year-ahead unemployment rate density forecasts}\label{app:unemp}
This appendix replicates the main inflation exercise for the ECB Survey of Professional Forecasters (SPF) \emph{unemployment rate} density forecasts. The goal is not to introduce a new methodology, but to verify whether the qualitative insights gained from the inflation study are applicable when our methodology is used for a different macroeconomic variable that possesses different time-series properties.

\subsection{Forecast object and target alignment}\label{app:unemp:setup}
Unemployment rate forecasts in the SPF are provided as individual density histograms over
pre-specified bins. We moment-match each forecaster's histogram to a Normal distribution
$\mathcal{N}(m_{t,j},s_{t,j})$ by using bin midpoints for the mean and adding within-bin variance
(Uniform-within-bin) to obtain $s_{t,j}$; this produces a comparable parametric representation across
time and forecasters while preserving the information in the elicited densities. As in the main text,
we record an availability indicator $i_{t,j}\in\{0,1\}$ for whether forecaster $j$ submits a density at
time $t$.

Unlike inflation (which in the SPF application is treated at a quarterly horizon), the unemployment rate
exercise uses the SPF {rolling} horizon convention: the ``one-year-ahead'' target is a specific
{calendar month} (e.g., \texttt{2026May}), whose definition depends on the latest available
unemployment rate observation at the time of the survey. For each survey quarter $t$, we therefore extract
the row corresponding to the SPF's one-year-ahead rolling \texttt{TARGET\_PERIOD} month and align the
realized unemployment rate to that month. The forecasting exercise is conducted at the one-year-ahead horizon:
point accuracy uses the predictive mean, and density accuracy uses the full predictive distribution.

\subsection{Forecaster participation and intermittency}\label{app:unemp:participation}
\begin{figure}[!t]
    \centering \setlength{\tabcolsep}{0\textwidth}
 \begin{tabular}{C{.5\textwidth}C{.5\textwidth}}   
    \includegraphics[width=0.49\textwidth]{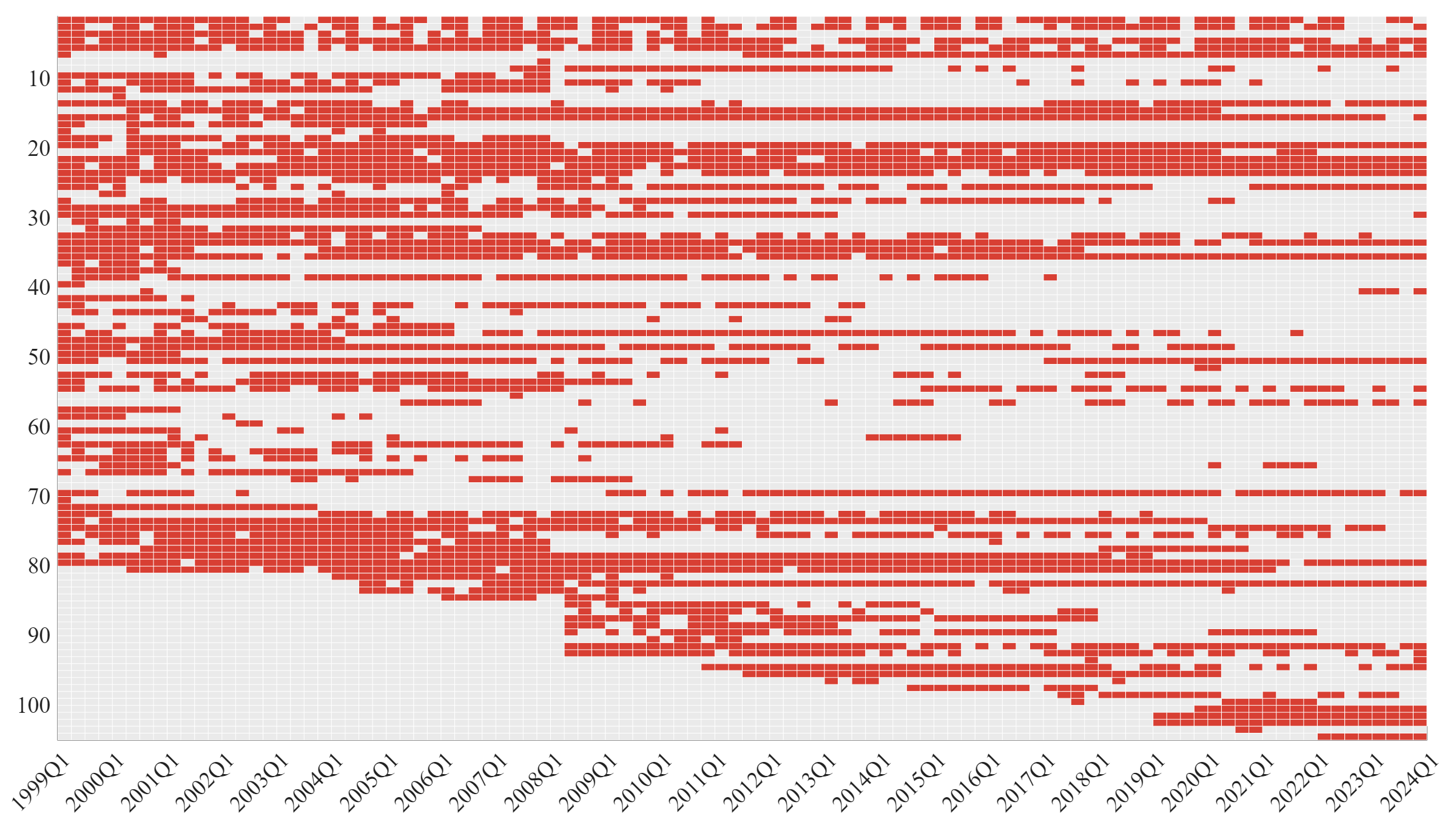}&    
    \includegraphics[width=0.49\textwidth]{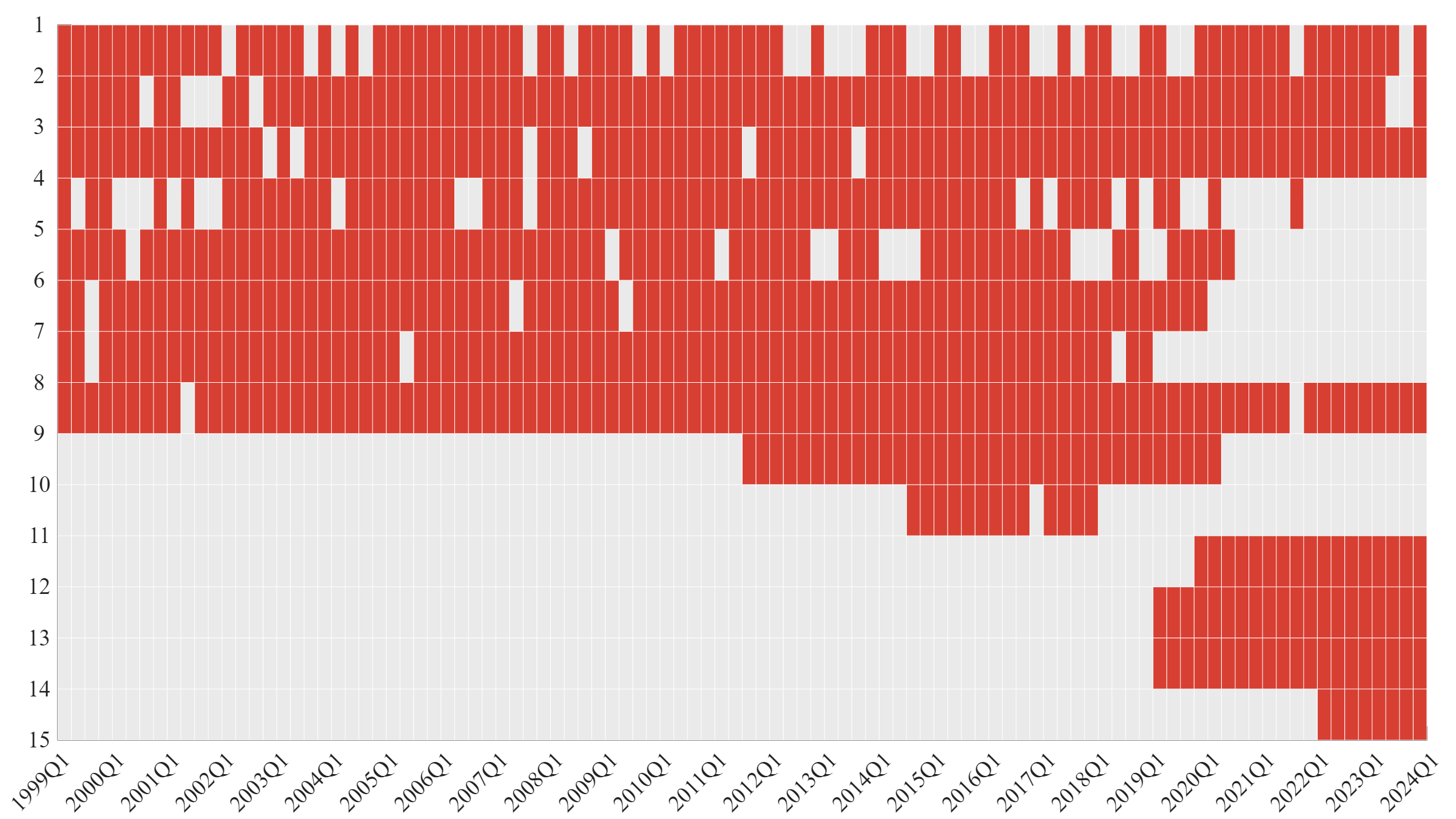}\\
    \small All forecasters &  \small Forecasters used in application \\
\end{tabular}
    \caption{Unemployment rate: Participation of expert forecasters over time. Left panel: participation of the full panel. Right panel: participation of the subset panel used for analysis.
    Light gray indicates inactivity and red denotes active contributions.}
    \label{fig:unemp_participation}
\end{figure}

Figure~\ref{fig:unemp_participation} summarizes participation for the unemployment rate densities.
Participation is strongly {sporadic}: there is a stable core of long-lived forecasters, but the
panel also exhibits extended gaps, short bursts of activity, and pronounced entry/exit episodes. In
particular, the later part of the sample contains substantial churn, with several forecasters entering
only after the mid-sample period and others exiting before the end. This pattern is qualitatively
similar to the inflation panel, but it is more severe because (i) the unemployment rate exhibits
larger blocks of missingness for some forecasters, and (ii) the post-2019 segment shows sharper entry
waves. These features make the unemployment rate exercise a useful stress test for the missingness mechanisms that motivate the synthesis framework.

\subsection{Competing combination rules}\label{app:unemp:methods}
We compare the same baselines and synthesis variants as in the inflation exercise. The operational
benchmark is the equal-weight (EW) mixture of available densities. Two common ``repair'' rules address
missingness prior to pooling: EW-LOCF replaces missing submissions by last-observation-carried-forward
densities; EW-ASMI uses agent-specific mean imputation. We also include an inverse-MSE scheme that
learns combination weights from past point forecast errors. The proposed approach applies BPS-style
synthesis to the collection of individual Normalized densities $\{\mathcal{N}(m_{t,j},s_{t,j})\}$ with
availability indicators $\{i_{t,j}\}$, using the same hyperparameter grid and $\theta^*$ choices as in
the main text: $\theta^*=0$ (no egalitarian pull), $\theta^*=1/J$ (weak shrinkage toward equal
weights), and $\theta^*=\text{prev}$ (temporal stabilization by shrinking toward the previous-period
weights). The persistence parameter $\rho$ controls the degree of discounting/adaptation.

\subsection{Full-sample performance}\label{app:unemp:results}
Table~\ref{tab:results_full_unemp} reports full-sample results over the evaluation period. The
main finding is that the unemployment rate is a case where synthesis delivers {large} gains in both
point and density performance relative to operational baselines.

\begin{table}[t]
\centering
\caption{Full-sample forecasting performance over the evaluation period.
RMSE measures one-year-ahead point accuracy; LPDR is the log predictive
density ratio against the equal-weight benchmark (positive is better). Target variable: unemployment rate. $J=16$}.
\small
\begin{tabular}{lllll}
 & EW & EW LOCF & EW ASMI & Inverse-MSE \\ \hline
RMSE & 1.7896 & 1.8702 & 2.4231 & 1.7844 \\
LPDR & - & -3.19 & 7.64 & 1.65 \\ \hline
 &  &  &  &  \\
 & \multicolumn{4}{c}{$\theta^*=0$} \\ \cline{2-5} 
 & $\rho=0$ & $\rho=0.5$ & $\rho=0.9$ & $\rho=0.99$ \\ \hline
RMSE & 1.1034 & 1.6549 & 1.7027 & 1.7025 \\
LPDR & 12.93 & 7.47 & 6.098 & 5.99 \\ \hline
 &  &  &  &  \\
 & \multicolumn{4}{c}{$\theta^*=1/J$} \\ \cline{2-5} 
 & $\rho=0$ & $\rho=0.5$ & $\rho=0.9$ & $\rho=0.99$ \\ \hline
RMSE & 1.1622 & 1.7052 & 1.7564 & 1.7451  \\
LPDR & 13.28 & 7.95 & 6.64 & 6.50 \\ \hline
 &  &  &  &  \\
 & \multicolumn{4}{c}{$\theta^*=prev$} \\ \cline{2-5} 
 & $\rho=0$ & $\rho=0.5$ & $\rho=0.9$ & $\rho=0.99$ \\ \hline
RMSE & 1.2446 & 1.7460 &  1.8081 & 1.9073 \\
LPDR & 12.84 & 6.92 & 5.57 & 5.44 \\ \hline
\end{tabular}

\label{tab:results_full_unemp}
\end{table}

First, consider point accuracy. The EW benchmark achieves an RMSE of $1.7896$, and inverse-MSE is
essentially indistinguishable ($1.7844$). By contrast, the best synthesis configuration substantially reduces RMSE, with RMSE as low as $1.1034$ (for $\theta^*=0$ and $\rho=0$) and materially 
below the baselines across the grid. The magnitude of this improvement is much larger than what is
typical in stable panels and suggests that the unemployment rate densities contain persistent,
forecaster-specific signal that is being diluted by naive pooling under entry/exit.

Second, density performance mirrors the point results. EW-LOCF is heavily penalized in LPDR
($-3.19$), consistent with the well known brittleness of LOCF when the underlying state changes.
EW-ASMI improves over EW in density terms (LPDR $7.64$) but performs poorly in RMSE, highlighting the
fact that ``fixing'' missingness in distribution space can distort the implied means. In contrast,
synthesis yields large positive LPDRs (roughly $5$--$13$ depending on $\rho$ and $\theta^*$), meaning
substantial cumulative predictive likelihood gains over EW.

The sensitivity patterns are also informative. When forecasting the unemployment rate, the best results occur at low $\rho$: $\rho=0$ dominates higher persistence settings for all $\theta^*$ choices in both RMSE and LPDR. This
suggests that the unemployment rate forecasting relationship is sufficiently time-varying that aggressive
discounting (fast adaptation) is beneficial. By comparison, the inflation exercise in the main text
exhibits weaker sensitivity and a more prominent role for stabilization (larger $\rho$ and/or
$\theta^*=\text{prev}$). In short, the unemployment appears to benefit more from rapid adaptation, whereas inflation benefits more from stabilization.

\subsection{Time-path diagnostics: why the gains arise}\label{app:unemp:diagnostics}
\begin{figure}[t]
    \centering
    \includegraphics[width=.95\linewidth]{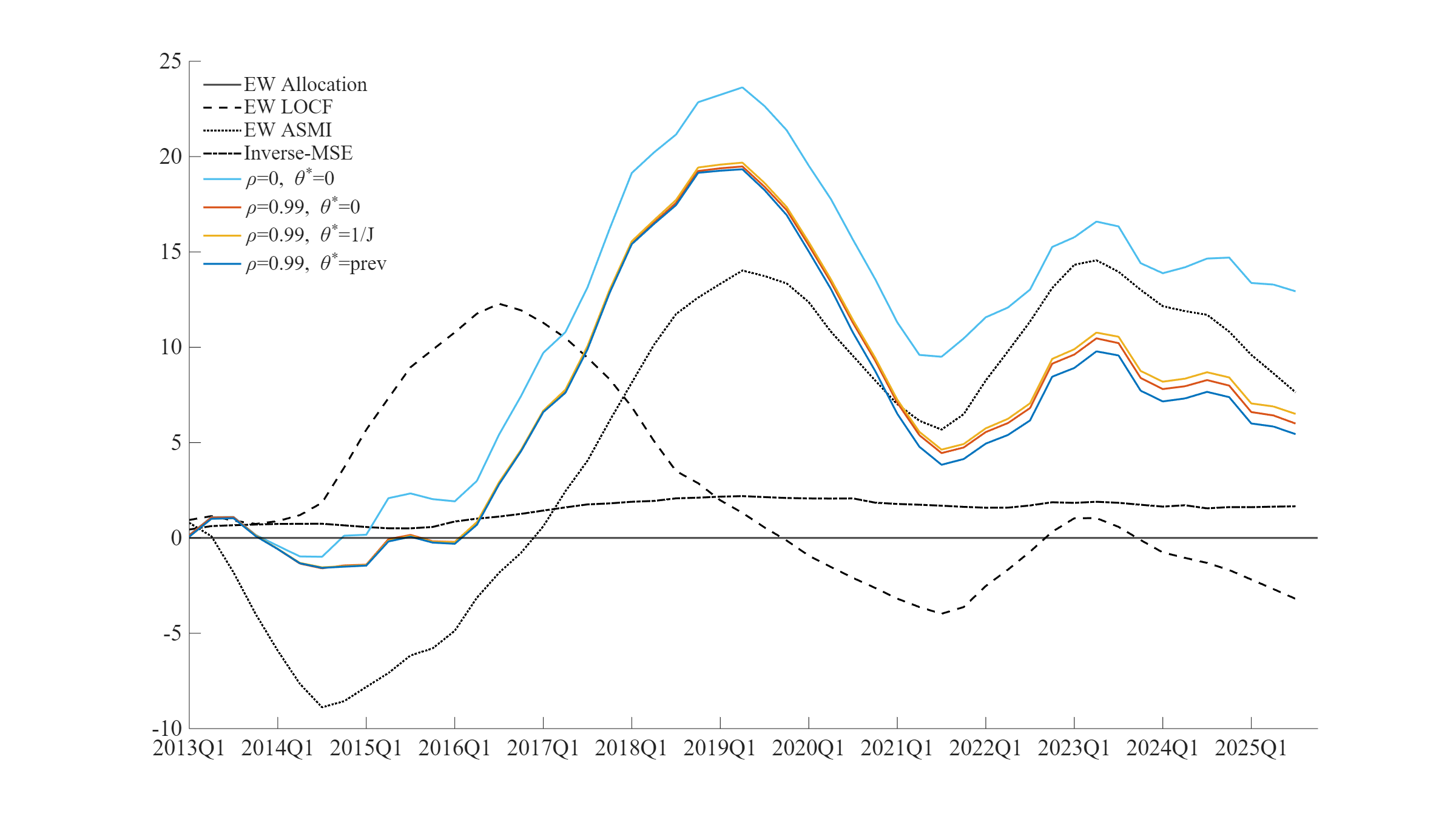}
    \caption{Cumulative LPDR (relative to EW) for unemployment rate one-year-ahead densities.}
    \label{fig:unemp_lpdr_path}
\end{figure}

\begin{figure}[t]
    \centering
    \includegraphics[width=.95\linewidth]{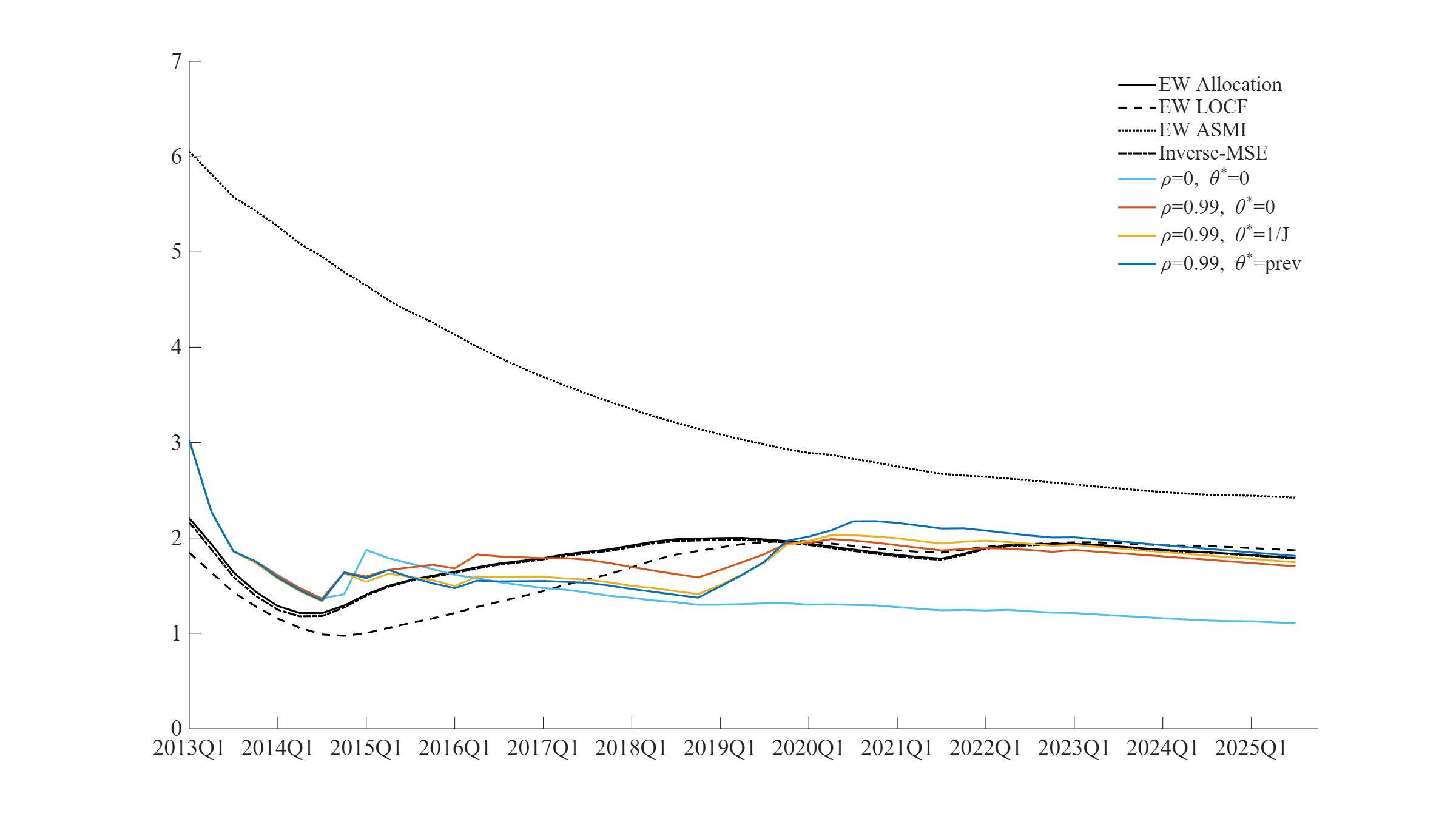}
    \caption{Running RMSE for unemployment rate one-year-ahead forecasts (point forecast = predictive mean).}
    \label{fig:unemp_rmse_path}
\end{figure}

Figures~\ref{fig:unemp_lpdr_path}--\ref{fig:unemp_rmse_path} plot the time paths of cumulative LPDR and
running RMSE, respectively. Two features stand out.

(i) The synthesis advantage is persistent rather than episodic. The cumulative LPDR curves for BPS
variants stay well above the baselines over long stretches, which indicates that improvements are not
driven solely by crisis quarters or one-off events.

(ii) LOCF fails precisely when adaptation matters. The LOCF curve underperforms markedly during
periods of trend changes and level shifts, consistent with imputing stale densities. This is the same
failure mode emphasized in the inflation exercise, but it is more pronounced here because the
unemployment rate is smoother and hence ``stale'' imputations are easier to diagnose in likelihood
terms.

\subsection{Contrast with the inflation exercise}\label{app:unemp:contrast}
Relative to the main inflation application, the unemployment rate results are stronger and more clear-cut.

\begin{itemize}
    \item \textbf{Bigger gains from synthesis.} In the unemployment rate exercise, synthesis reduces RMSE from roughly
    $1.8$ (EW) to about $1.1$ at the best setting, and delivers double-digit LPDR improvements. The
    inflation exercise shows improvements as well, but typically of smaller magnitude, consistent with
    inflation being harder to forecast and more sensitive to common shocks that affect all agents.
    
    \item \textbf{Different optimal adaptivity.} The unemployment rate favors low $\rho$ (rapid adaptation),
    whereas inflation tends to benefit more from persistent structure (higher $\rho$ and/or
    $\theta^*=\text{prev}$). This difference is plausible: unemployment evolves more smoothly but its
    forecast mapping can change with regime/policy; inflation exhibits sharper shocks where
    stabilization prevents overreaction to short-lived forecast anomalies.
    
    \item \textbf{Baseline behavior differs under imputation.} In the unemployment rate exercise, EW-ASMI attains a sizable LPDR yet worsens RMSE, illustrating a disconnect between distribution-space imputation and
    mean accuracy. In the inflation exercise, the same baselines are typically more aligned, implying
    that the mean/shape tradeoff is more severe for unemployment densities (likely due to tighter bin
    structure and smaller intrinsic volatility).
\end{itemize}

Overall, the unemployment rate exercise reinforces the main message: when forecaster participation is
sporadic and the panel exhibits entry/exit, synthesis methods that explicitly account for missingness
and learn time-varying combination structure can dominate both operational EW pooling and common
imputation-and-average heuristics.

\end{document}